 \documentclass[final,5p,times,twocolumn,fleqn]{elsarticle}

\usepackage{amssymb}
\usepackage{amsmath,amsthm,amssymb,cases}
\usepackage[linesnumbered,ruled]{algorithm2e}
\usepackage{hyperref}
\usepackage{graphicx}
\usepackage{comment}
\usepackage{subfig}
\usepackage{array}
\usepackage{picins}
\usepackage{url}
\usepackage{color}
\usepackage{here}

\theoremstyle{definition}
\newtheorem{definition}{Definition}[section]
\theoremstyle{remark}

\newcommand{\xmark}{\ding{55}}%

\newcommand{\mymin}{\mathop{\rm min}\limits}
\newcommand{\mymax}{\mathop{\rm max}\limits}
\newcommand{\mysum}{\mathop{\rm \sum}\limits}

\newcommand\boldred[1]{#1}

\journal{Journal of Parallel and Distributed Computing (JPDC), Elsevier, 2017}

\begin{document}

\begin{frontmatter}



\title{Scheduling Parallel and Distributed Processing \\for Automotive Data Stream Management System}


\author[RG]{Jaeyong Rho}
\author[O]{Takuya Azumi}
\author[R]{Mayo Nakagawa}
\author[D]{Kenya Sato}
\author[R]{Nobuhiko Nishio}

\address[RG]{Graduate School of Information Science and Engineering, Ritsumeikan University, Kusatsu, Shiga 5258577, Japan}
\address[R]{College of Information Science and Engineering, Ritsumeikan University, Kusatsu, Shiga 5258577, Japan}
\address[O]{Graduate School of Engineering Science, Osaka University, Toyonaka, Osaka 5608531, Japan}
\address[D]{Mobility Research Center, Doshisha University, Kyoto 6028580, Japan}

%\address[1]{no@ubi.cs.ritsumei.ac.jp}
%\address[2]{takuya@sys.es.osaka-u.ac.jp}
%\address[3]{mayo@ubi.cs.ritsumei.ac.jp}
%\address[4]{ksato@mail.doshisha.ac.jp}
%\address[5]{nishio@cs.ritsumei.ac.jp}

\begin{abstract}
In this paper, \boldred{to analyze end-to-end timing behavior in heterogeneous processor and network environments accurately, we adopt and modify a heterogeneous selection value on communication contention (HSV\_CC) algorithm, which can synchronize tasks and messages simultaneously, for stream processing distribution.} \boldred{In order to adapt the concepts of a static algorithm like HSV\_CC to automotive data stream management system (DSMSs), one must first address three issues: (i) previous task and message schedules might lead to less efficient resource usages in this scenario; (ii) the conventional method to determine the task scheduling order may not be best suited to deal with stream processing graphs, and; (iii) there is a need to be able to schedule tasks with time-varying computational requirements efficiently.} To address (i), we propose the heterogeneous value with load balancing and communication contention (HVLB\_CC) (A) algorithm, which considers load balancing in addition to the parameters considered by the HSV\_CC algorithm. We propose HVLB\_CC (B) to address issue (ii). HVLB\_CC (B) can deal with stream processing task graphs and more various directed acyclic graphs to prevent assigning a higher priority to successor tasks. In addition, to address issue (iii), we propose HVLB\_CC\_IC. To schedule tasks more efficiently with various computation times, HVLB\_CC\_IC utilizes schedule holes left in processors. These idle time slots can be used for the execution of an optional part to generate more precise data results by applying imprecise computation models. Experimental results demonstrate that the proposed algorithms improve minimum schedule length, accuracy, and load balancing significantly compared to the HSV\_CC algorithm. In addition, the proposed HVLB\_CC (B) algorithm can schedule more varied task graphs without reducing performance, and, using imprecise computation models, HVLB\_CC\_IC yields higher precision data than HVLB\_CC without imprecise computation models.

\end{abstract}


\end{frontmatter}


\section{Introduction}\label{sec:motivation_2}
Modern automotive systems incorporate a range of data from on-board and external sensors. Advanced driver-assistance systems, such as collision 
warning systems \cite{its}, help prevent accidents by alerting drivers to oncoming 
vehicles or pedestrians, which requires monitoring of the environment. Autonomous driving systems, such as the Google 
driverless car, should be able to determine when to change lanes and 
 take appropriate action to avoid obstacles. To realize these functionalities, contemporary automotive systems require many types of data, such as own-vehicle 
 and surrounding-vehicle information.

Automotive systems are distributed systems. Currently, modern luxury vehicles have more than 70 electronic control units (ECUs) connected to various in-vehicle networks \cite{70ecu}. These ECUs comprise different types of microcontrollers, from simple 4-bit controllers to complex 32-bit controllers, to handle different operations \cite{diff_ecu}. Various types of microcontrollers are managed by ECUs that are connected via multiple in-vehicle networks, such as local interconnect networks (LINs), controller area networks (CANs), and media-oriented system transport (MOST) networks \cite{in_vehicle_networks}. Thus, automotive software development is complex, and as the complexity of data processing and amount of data increases, automotive software development will become increasingly complex.

Research has been conducted on the adaptation of data stream management systems (DSMSs) for automotive embedded systems to reduce data processing complexity and lower software 
development costs while increasing the amount of available data \cite{crts}. In current systems, automotive data from on-board and external sensors are processed and 
managed individually by separate electronic control units (ECUs). The duplication of data processing over multiple 
applications and the associated software development costs increase with increased amounts of data. Thus, researchers have focused on DSMSs for automotive system applications that can process streamed data 
at low latency using processes that are shared over multiple applications \cite{crts,aedsms,streamcar}. 

Existing DSMSs \cite{borealis,aurora,telegraph,stream} are designed to run on general-purpose computers and primarily target network monitoring and financial analysis applications. Such applications require abundant resources but do not necessarily consider strict real-time constraints. In general-purpose DSMSs, dynamic distribution and stream processing optimization at runtime over multiple processors \cite{multicore_dsms,multicore_dsms2,multicore_dsms3} are commonly adopted. These dynamic methods are difficult to implement in automotive embedded systems because time predictability for stream processing cannot be guaranteed and because a large number of modules would be required, which is not appropriate for automotive systems. Therefore, a distribution method that can guarantee strict time predictability for stream processing is required. 

In distributed real-time systems, application tasks usually have precedence constraints so that the results can be used as input data for other tasks. Real-time applications can be described as directed acyclic graphs (DAGs) that have end-to-end deadlines in order to take advantage of parallel distributed processing. Parallel processing performance is highly dependent on task and message scheduling \cite{hetero_computing1,hetero_computing2}. Scheduling distributes multiple tasks and messages to realize parallel processing and to satisfy precedence requirements. Scheduling an application on multiple processors to minimize overall scheduling length and parallel processing time is a multi-processor scheduling problem, which is recognized as an NP-complete optimization problem. Thus, heuristics are utilized to obtain near-optimal solutions than searching all possible scheduling patterns, which is not possible for realistic large-scale scheduling problems.

Most heuristic scheduling algorithms are based on list scheduling \cite{list1,list2,list3,list4,list5,clus5,list7}. List scheduling consists of two phases: (i) a task prioritizing phase in which tasks in a DAG are listed by priority, and (ii) a processor-selection phase, in which the tasks with the highest priorities are scheduled. List scheduling is a generally accepted method because it provides high-quality scheduling and has low complexity \cite{list_advantage}. Thus, many list scheduling algorithms have been proposed for near-optimal solutions in parallel and distributed systems.

However, achieving time accuracy in heterogeneous environments is difficult for most list scheduling algorithms because they use the average computation time for each task on heterogeneous processors, even though different computation times must be considered. Furthermore, they assume that all processors are fully connected so that communication can occur simultaneously. This can lead to inaccurate scheduling results in automotive systems. Given these ideal assumptions of existing list scheduling algorithms, such algorithms have not been widely utilized in heterogeneous computing systems. 

To address these assumptions, Xie et al. \cite{hsv_cc} proposed heterogenous selection value on communication contention (HSV\_CC) for automotive embedded systems based on list scheduling. HSV\_CC can schedule tasks and messages simultaneously and considers communication contention. However, HSV\_CC has several issues when applied to the automotive DSMSs. First, task and message scheduling results can lead to inefficient resource usage. Tasks and messages are likely to be assigned to specific processors and network links, and, in the worst case, some processors and links are poorly utilized. \boldred{Second, HSV\_CC was designed to schedule DAGs and consequently, some stream processing applications for automotive DSMSs may not be scheduled appropriately.} Third, \boldred{tasks whose requested processing times can vary over time have to be scheduled efficiently to apply HSV\_CC to automotive DSMSs.}
%{ \color{black}Thus, HSV\_CC cannot utilize resources sufficiently and schedule DAGs in parallel.} Second, HSV\_CC scheduling is restricted to various DAGs. Consequently, some stream processing applications for automotive DSMSs will not be scheduled. Third, HSV\_CC cannot be applied to stream processing if the request processing time can vary with the time.

We propose the heterogeneous value with load balancing and communication contention (HVLB\_CC) static list scheduling algorithm, which considers load balancing (LB) over multiple processors and can be applied to automotive DSMSs. When scheduling tasks and messages, both the HSV\_CC parameters and parameters that balance the load among the processors are considered. In addition, this method achieves schedule length reduction by considering processor loads to utilize resources more efficiently. Our main contributions are as follows.

$\bullet$ We propose a list scheduling algorithm that considers processor LB to reduce parallel execution times (HVLB\_CC (A)). The proposed algorithm can deal with important load balancing parameters in addition to HSV\_CC parameters. We show that HVLB\_CC can produce more accurate scheduling results with task and message synchronization. HVLB\_CC can also be applied to real-time heterogeneous distributed computing systems.

$\bullet$ We address the scheduling problem that occurs in HSV\_CC, i.e., when scheduling task graphs for stream processing, tasks are scheduled before the predecessor tasks are scheduled (HVLB\_CC (B)). Compared to HSV\_CC, the proposed algorithm can schedule stream processing graphs (SPGs) and different types of DAGs.

$\bullet$ We extend the HVLB\_CC algorithm to execute data streams more efficiently (HVLB\_CC\_IC). HVLB\_CC can exploit schedule holes to deal with varying requested processing times. We utilize imprecise computations for tasks with varying processing times. This approach produces more precise data than an approach that do not utilize imprecise computation models can produce.

$\bullet$ Compared to HSV\_CC, the proposed algorithm demonstrates improved {\it speedup}, scheduling accuracy, and LB. Furthermore, the extended version of HVLB\_CC with imprecise computations can produce more precise data results than the original can produce.

The remainder of this paper is organized as follows. A system model and a stream processing application model are described in Section \ref{sec:system_model2}. The problem is defined in Section \ref{sec:problem2}. Section \ref{sec:algorithm2} details the proposed algorithm. Simulation
results are presented in Section \ref{sec:simulation2}. In Section \ref{sec:related_work2}, previous work related to scheduling algorithms on heterogeneous distributed systems is reviewed. Conclusions and suggestions for future work are given in Section \ref{sec:conclusion2}.

%%%%%%%%%%%%%%%%%%%%%%%%%%%%%%%%%%%%%%%%%%%%%%%%%%%%%%%%%%%%%%%%%%%%

\section{System Model}\label{sec:system_model2}
In this section, we describe a DSMS for automotive embedded systems \cite{aedsms3,aedsms2}, assumptions about the target system, and a stream processing application model.

\subsection{Stream Processing}
\begin{figure}
\centering
\includegraphics[width=1.01\linewidth]{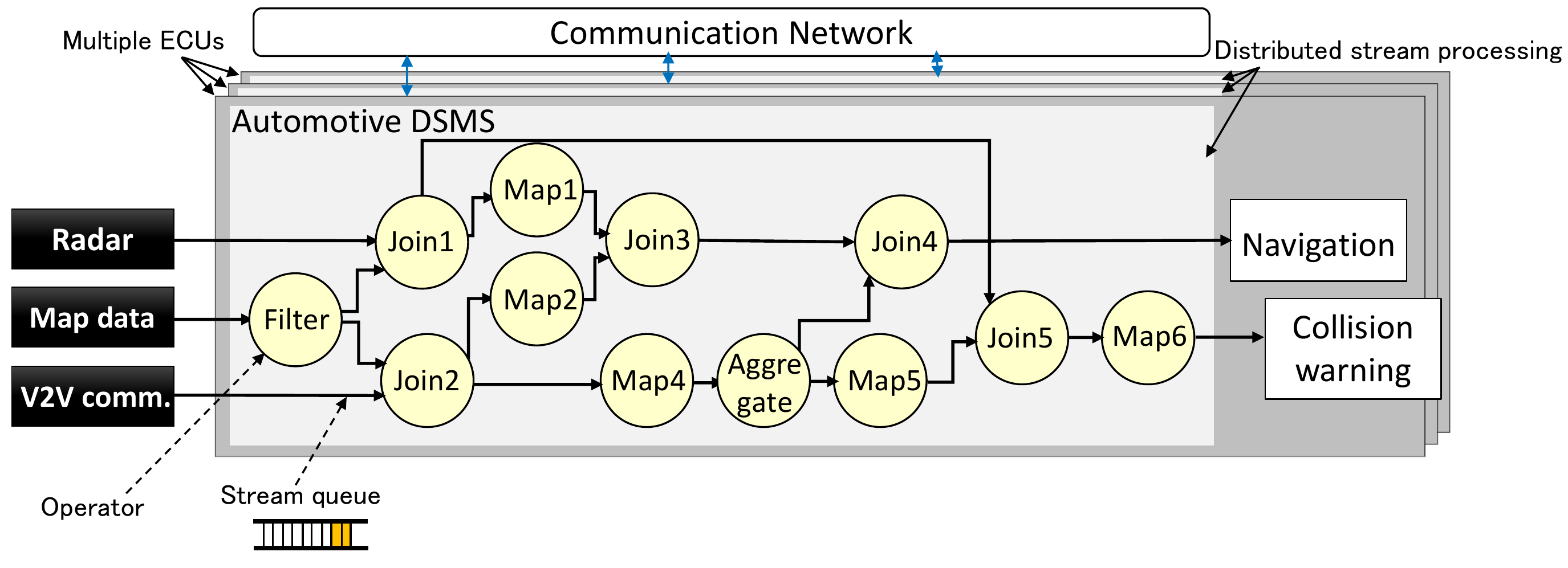}
\caption{Adapting DSMS to automotive systems and sample query for stream processing}
\label{fig:entire_dsms}
\vspace{-0.3cm}
\end{figure}

A DSMS processes {\it stream data} generated continuously in large volumes in real-time using a query. The query is registered with the system and executes 
continuously as new stream data arrives. By registering a query in advance, the overhead required to issue a query for each process is smaller than that required in a DBMS. This enables processing in real time. A {\it query} consists of one or 
more operators, and two consecutive operators are connected with a {\it stream queue}. An {\it operator} is a communication entity that uses a tuple, i.e., a set of data values, and outputs the computed result tuples to the next stream queue. For example, in \textbf{Figure} \ref{fig:entire_dsms}, the circle and arrow in the query file represent an operator and a stream queue, respectively.

In stream processing, a precedent 
operator executes and produces one or more result tuples using input 
tuples. The result tuples are delivered to subsequent operators or applications through a stream queue. Operators used for automotive DSMSs, such as Map, Filter, and Join, are based on Borealis, which is a general-purpose DSMS \cite{borealis}. The operator functionalities and other details are described in \cite{aedsms,aedsms3}.

There are two main differences between automotive and general DSMSs. The first difference is that the automotive DSMS requires synchronous execution. The in-vehicle system has different data acquisition rates for the in-vehicle sensors and the outside vehicle communication. Therefore, the timestamps need to be aligned to synchronize execution. Second, static scheduling needs to be implemented in the in-vehicle system because it is not necessary to guarantee strict real-time processing in this case.

\subsection{DSMS-Based Data Integrated Architecture}
In the current automotive system architecture, data retrieved from on-board and external sensors are 
processed individually by separate applications embedded in an ECU because suppliers provide a product as a set consisting of software and 
the related sensors. Therefore, data processes can be duplicated 
over multiple applications in different ECUs. Furthermore, when system properties (e.g., sensor type) must be changed, large parts of the application 
programs must be modified. To tackle these issues, automotive DSMSs have been 
developed based on the data integrated architecture \cite{aedsms,aedsms3} shown in \textbf{Figure} \ref{fig:entire_dsms}. Applications can be 
separated from sensor devices because data processing for multiple 
sensors is defined in the DSMS rather than in the sensors. Thus, data from multiple applications is accessible in a location-transparent manner. In addition, changing system properties becomes easier because applications are independent from specific sensors; thus, automotive software development costs can be reduced.

\subsection{Heterogeneous Processors and Networks}
Each operator (task) has different computation times on different ECUs (processors), because different ECUs have different capabilities. Each node $n_i$ in a DAG has a weight $w_i$ that represents its computational volume, i.e., the amount of computation operations needed to be executed. Each processor $p_{u}$ is characterized by an execution rate $\mu_{p_u}$ because the capabilities of heterogeneous processors differ. 

\theoremstyle{definition}
\begin{definition}{Computation time on heterogeneous processors. The processor execution rates vary according to the processor capabilities. The computation time of task $n_i$ on $p_u$ is given by:}\label{def:comp_time}
\begin{equation}
\label{eq:comp_time}
\hspace*{-7mm} comp({n_i,p_{u}})= \frac{w_i}{\mu_{p_u}},
\end{equation}
\end{definition}
where $\mu_{p_u}$ is the execution rate of processor $p_{u}$. 

For example, assume that the weight of $n_6$ is 10 and that the execution rates of processors $p_1$, $p_2$, and $p_3$ are 0.67, 1.0, and 0.83, respectively. Then, the computation times of $n_6$ on processors $p_1$, $p_2$, and $p_3$ are 15, 10, and 12, respectively. Computation times for other tasks can be calculated in the same manner, as shown in \textbf{Table} \ref{tab:comp_time}. 

There are various types of network topologies in automotive networks, including bus, star, ring, tree, and mesh types. For example, CANs and FlexRay are configured with bus topologies; however, they can be divided by gateways to form other topologies over different domains. MOST is typically configured with a ring topology. Thus, automotive networks have a hybrid network topologies that consist of different network technologies with different bandwidths. Therefore, identical messages transmitted by different links have different communication speeds.

\begin{table}[t]
\centering
\caption{Computation time matrix corresponding to Fig \ref{fig:dsg_sample}}
\label{tab:comp_time}
%\footnotesize
\scriptsize
\begin{tabular}{p{1.5cm}p{1.3cm}p{1.3cm}p{0.7cm}} \hline
Task&$p_1$&$p_2$&$p_3$\\[0.7ex] \hline
$n_1$&18&12&14\\
$n_2$&12&8&10\\
$n_3$&12&8&10\\
$n_4$&21&14&17\\
$n_5$&9&6&7\\
$n_6$&15&10&12\\
$n_7$&26&17&20\\
$n_8$&14&9&11\\
$n_9$&20&13&16\\
$n_{10}$&15&10&12\\
\hline\end{tabular}
\vspace{-0.3cm}
\end{table}

\begin{figure}[!t]
\centering
\includegraphics[width=0.58\linewidth]{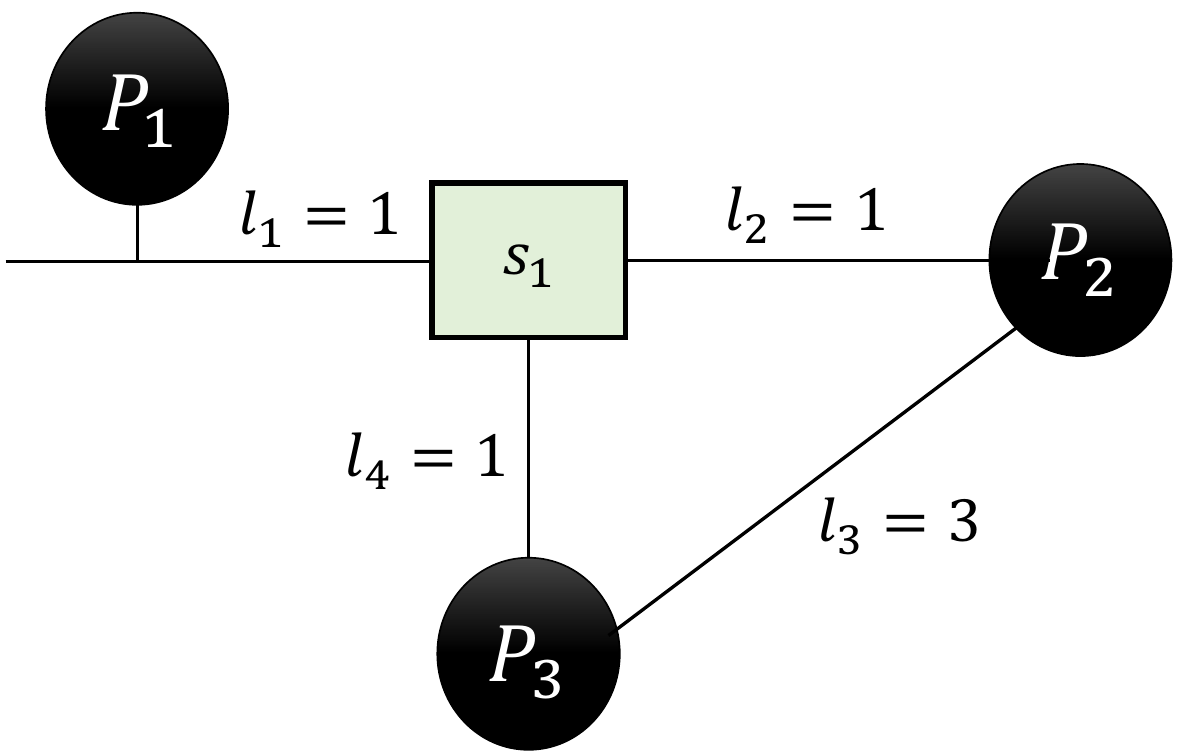}
\caption{Example of heterogeneous network topology}
\label{fig:network_topology}
\vspace{-0.3cm}
\end{figure}

A heterogeneous network can be described using an undirected graph $TG=\langle P,S,L\rangle$. {\it P} is a set of heterogeneous processors, and {\it S} is a set of switches or gateways. {\it L} represents a set of {\it v} links, $L=\{l_1,l_2,...,l_v\}$, and each link value represents its communication speed. A message can be expressed as $e_{i,j}^{P_{src},P_{dest}}$ and is transmitted from task $n_i$ to task $n_j$. $P_{src}$ is a source processor assigned to $n_i$ that sends the message to the immediate successor task, and $P_{dest}$ is a destination processor assigned to $n_j$ that receives the message from the predecessor task. There are one or more routes between $P_{src}$ and $P_{dest}$ to transmit the message $e_{i,j}^{P_{src},P_{dest}}$. The {\it z} routes between two processors can be expressed as $R^{P_{src},P_{dest}}=\{r_1^{P_{src},P_{dest}},r_2^{P_{src},P_{dest}},...,r_z^{P_{src},P_{dest}}\}$. Each route set in $R^{P_{src},P_{dest}}$ consists of one or more links with different communication speeds.

An example heterogeneous automotive network topology is shown in \textbf{Figure} \ref{fig:network_topology} \cite{hsv_cc}. There are two routes between $P_1$ and $P_3$, denoted $R^{P_{1},P_{3}}=\{r_1^{P_{1},P_{3}}, r_2^{P_{1},P_{3}}\}$. The route sets are expressed as $r_1^{P_{1},P_{3}}=\{l_1^1,l_4^1\}$ and $r_2^{P_{1},P_{3}}=\{l_1^2,l_2^2,l_3^2\}$.

\begin{figure}
\includegraphics[width=1.2\linewidth]{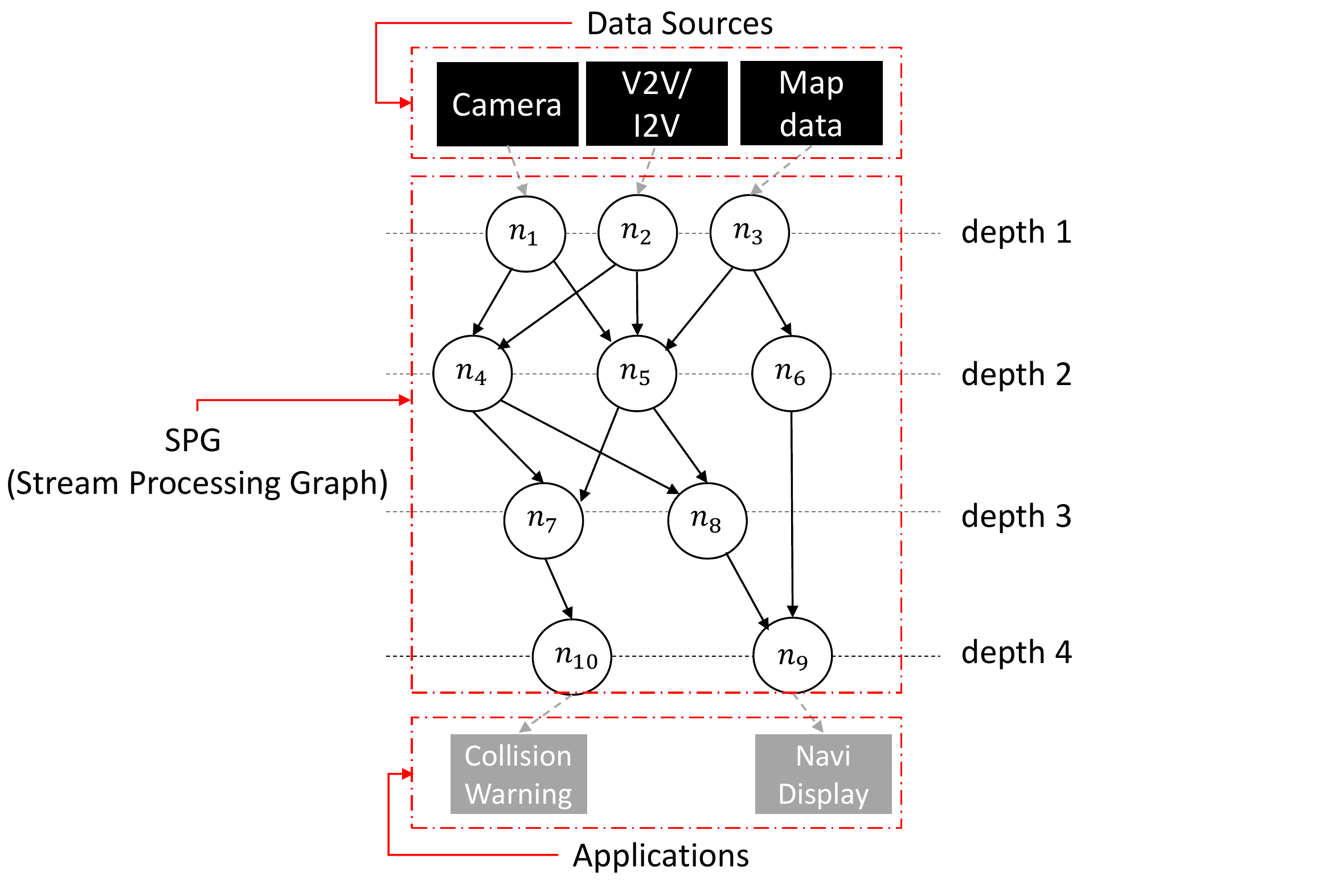}
\caption{Example of SPG}
\label{fig:dsg_sample}
\vspace{-0.3cm}
\end{figure}

\subsection{Stream Processing Application}
A stream processing application can be represented by a SPG, as shown in Figure 3, and by a DAG; thus, an SPG for a stream processing application is a type of DAG. An SPG can have tasks with a greater out- degree than the out-degree of its predecessor.

\theoremstyle{definition}
\begin{definition}{Stream Processing Graph. An SPG is a DAG consisting of a node set and a directed edge set. It can be denoted $G=\langle V(G),E(G)\rangle$, where $V(G)=\{n_1,n_2,...,n_j\}$ is a finite set of $j$ nodes in the graph G that represents tasks, and $E(G)=\{e_{1,2},e_{1,3},...,e_{n-i,n}\}$ is a finite set of directed edges that represents execution precedence between nodes.}
\end{definition}

In the SPG shown in \textbf{Figure} \ref{fig:dsg_sample}, there is a set of 10 nodes $V(G)=\{n_1,n_2,...,n_{10}\}$. Hereafter, the terms {\it node} and {\it task} are used interchangeably. We assume that a task can be executed only if all the predecessor operators have completed execution because the rate of data acquisition from each information source in the vehicle DSMS is different. When the data time stamps do not match in synchronous execution, incorrect results may be obtained.
 {\it E} is a set of directed edges $e_{i,j}\in E$ that indicates data dependencies between tasks $n_i$ and $n_j$, representing that task $n_j$ can begin execution only after task $n_i$ completes transmission of the result. Each edge $e_{i,j}\in E$ has a weight ${\it tpl}(e_{i,j})$ that represents its communication volume, i.e., the amount of data (a tuple) to be transferred from task $n_i$ to $n_j$.

The source node of an edge is called the $predecessor$ and the sink node of the edge is called the {\it successor}. $pred(n_i)$ denotes the set of $n_i$'s immediate predecessor tasks. $ind(n_i)$ represents $n_i$'s in-degree, i.e., the cardinality of $pred(n_i)$. For example, in \textbf{Figure} \ref{fig:dsg_sample}, $pred(n_5)=\{n_1,n_2,n_3\}$ and $ind(n_5)=3$. We assume that a task can be executed only if all of predecessor operators have completed execution. $succ(n_i)$ is the set of $n_i$'s immediate successor tasks and $outd(n_i)$ is $n_i$'s out-degree, which indicates the cardinality of $succ(n_i)$. For example, $succ(n_5)=\{n_7,n_8\}$ and $outd(n_5)=2$ in \textbf{Figure} \ref{fig:dsg_sample}. A task with no predecessor task is called an {\it entry} task, $n_{entry}$, and a task with no successor is called an {\it exit} task, $n_{exit}$. 

Each SPG has an end-to-end deadline. The tasks in the SPG do not have individual deadlines. Therefore, when all {\it exit} tasks finish execution before the DAG's end-to-end deadline, the DAG can guarantee timing constraints. The {\it depth} of a task is the length of the longest path from an {\it entry} task to the particular task. The {\it depth} starts at 1; thus, the {\it depth} of an {\it entry} task is 1. The {\it length} of the path in the graph is equal to the number of edges on the path. For example, the longest path of $n_9$ is \{$n_3,n_5,n_8,n_9$\}, and its {\it length} is 3 because there are three edges \{$e_{3,5}$,$e_{5,8}$,$e_{8,9}$\}. Thus, the {\it depth} of $n_9$ is $1+3=4$. 

Sources that generate sensor data (tuples) to an automotive DSMS can be on-board sensors such as cameras, radar, GPS systems, and external communication systems, such as vehicle-to-vehicle (V2V) and infrastructure-to-vehicle (I2V) communication systems. Various applications use stream processing results, e.g., collision warnings, automotive navigation systems, and vehicle-infrastructure cooperative right-turn collision caution signals.

%%%%%%%%%%%%%%%%%%%%%%%%%%%%%%%%%%%%%%%%%%%%%%%%%%%%%%%%%%%%%%%%%%

\section{Problem Description}\label{sec:problem2}
\subsection{Inefficient Resource Usage}
To achieve high performance in distributed and parallel stream processing, it is essential to maintain a well-balanced load among all available resources, such as processors and communication links. The previously proposed HSV\_CC \cite{hsv_cc} may result in inefficient resource utilization. In the processor selection phase, each task is assigned to a processor where the minimum {\it HSV\_CC} value can be obtained. The {\it HSV\_CC} value can be calculated by using ${\it HSV\_CC}(n_j,p_{dest},r_z^{p_{src},p_{dest}})=EFT(n_j,p_{dest},r_z^{p_{src},p_{dest}}) \times LDET\_CC(n_i,p_{dest})$, where {\it EFT} is the earliest finish time and {\it LDET\_CC} is the longest distance exit time. {\it LDET\_CC} is calculated by using $rank(n_i,p_u)-comp(n_i,p_u)$. $rank(n_i,p_u)$ is calculated by using $comp$ and $comm$; thus, the {\it rank} value will be smaller for higher-capability processors and communication links. Similarly, {\it EFT} will be small for each task. Therefore, the factors required to calculate {\it HSV\_CC} value are highly dependent on the processor and communication link capabilities. Inefficient scheduling may occur because tasks and messages tend to be assigned to high-capability processors and communication links. 

We describe a scheduling example in which tasks and messages are assigned to processors connected by links with different capabilities. As a task set, we use the task graph shown in \textbf{Figure} \ref{fig:dsg_sample} and the network topology shown in \textbf{Figure} \ref{fig:network_topology}. Here, there are three processors ($p_1$, $p_2$, and $p_3$), and the computation times for 10 tasks are given in \textbf{Table} \ref{tab:comp_time}. The execution rates of the three processors are 0.67, 1.0, and 0.83.

\begin{figure}[t]
\centering
\includegraphics[width=1.01\linewidth]{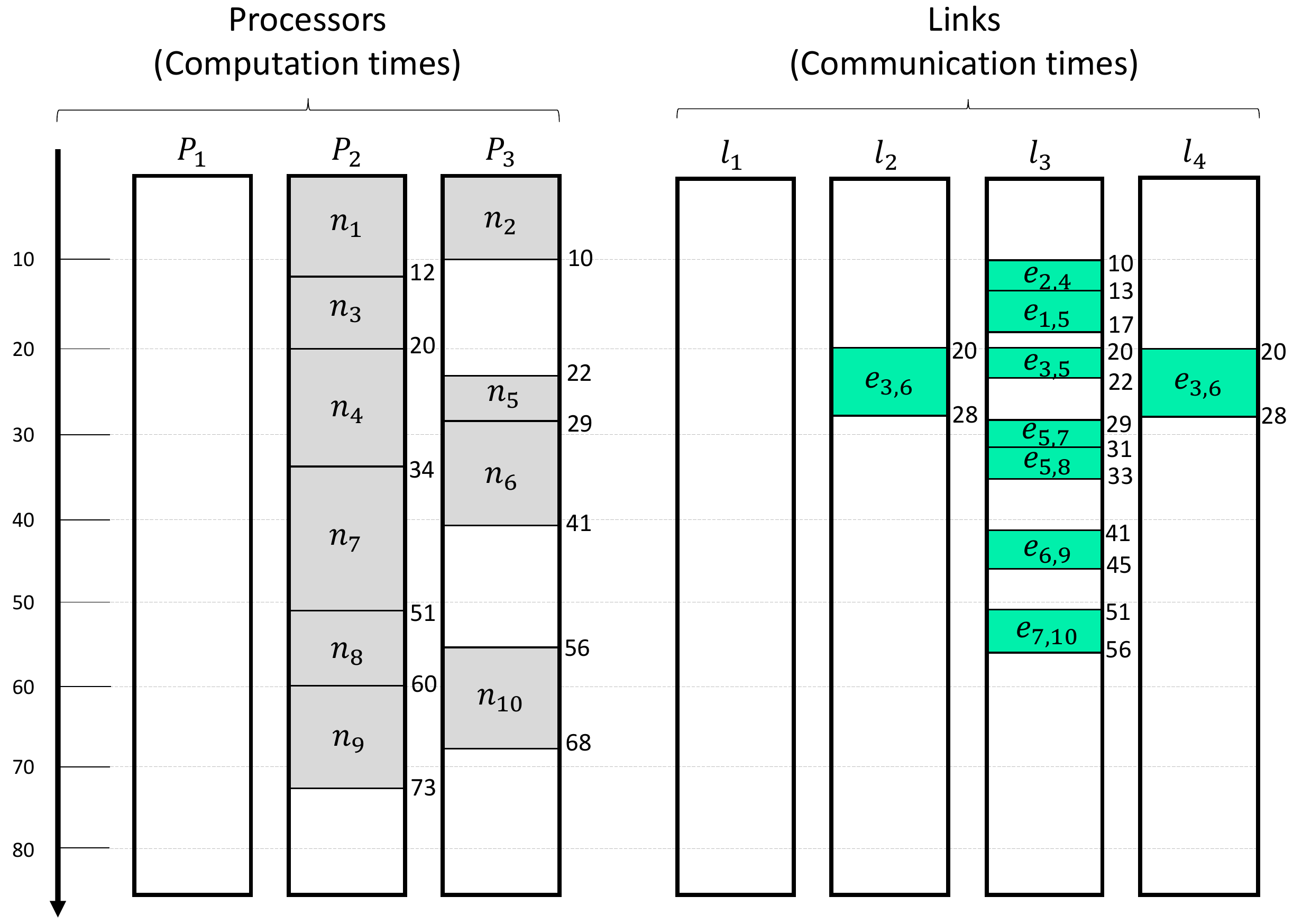}
\caption{Gantt chart for SPG scheduling in Fig. \ref{fig:dsg_sample} with HSV\_CC algorithm ({\it makespan}=73)}
\label{fig:hsv_cc_assign}
\vspace{-0.3cm}
\end{figure}

The scheduling results of HSV\_CC are shown in \textbf{Figure} \ref{fig:hsv_cc_assign}. Six tasks are assigned to $p_2$, and four tasks are assigned to $p_3$. No tasks are assigned to $p_1$. Communication link $l_1$ is not used because no messages are transmitted to $p_1$. Furthermore, $l_2$ and $l_4$ only transmit messages from task $n_3$ to task $n_6$, and all other messages are transmitted through link $l_3$ even though $l_2$ and $l_4$ are not busy. This allocation pattern for tasks and messages, as scheduled by HSV\_CC, is difficult to deal with in a situation where the amount of input sensor data and sensor fusion processing increases rapidly, because there is limited idle time after finishing a task processing and transmitting a message on the processor and communication link as most tasks and messages are concentrated on specific resources. Thus, it is evident that allocation of tasks and messages to processors and links with HSV\_CC can result in inefficient resource usage.

\subsection{Scheduling the Stream Processing Graph}
In HSV\_CC, a task cannot be scheduled before its predecessor has been scheduled because the task computes {\it EFT}, which is used to select an appropriate processor based on the actual finish time of the predecessor. Assume that $n_j$ whose $outd(n_j)=3$ uses the results generated from $n_i$ whose $outd(n_i)=1$. In this case, it is a possible that $n_j$ could be assigned a higher priority than $n_i$ is assigned, and $n_j$ may be scheduled before its predecessor task has completed because, in HSV\_CC, task priority is determined by $HPRV\_CC(n_i)=hrank(n_i) \times outd(n_i)$. Thus, a node with a large out-degree value can have a higher $HPRV\_CC$ value than its predecessor if the gap between their {\it hrank} values, which indicates the average {\it rank} value over all processors, is not significant. 
In an automotive DSMS, when stream processing must be shared among multiple applications (sink nodes), then a task located at high depth in a task graph can have a higher out-degree than the lower-depth tasks have. Thus, a high-depth task may have a higher priority than its predecessors.

\subsection{Stream Processing with Varying Execution Times}
A scheduling approach for automotive DSMSs should deal with variable execution times because the arrival rate of the input data can be increased. For example, in \textbf{Figure} \ref{fig:dsg_sample}, it is assumed that $n_2$ inputs stream data from an external source, such as V2V and I2V, and transforms the data format. Here, $n_5$ is a matching process with map data and external vehicle information. The arrival rates of input data to $n_2$ and $n_5$ can vary because the volumes and arrival times of the data streams from external sources can change depending on the environmental conditions, e.g., many V2V data streams are generated at congested urban intersections. However, static list scheduling HSV\_CC considers tasks whose computation times do not vary. Therefore, a task with varying computation times is difficult to be scheduled efficiently, and data quality can decrease as the arrival rate of an external data stream increases.

%%%%%%%%%%%%%%%%%%%%%%%%%%%%%%%%%%%%%%%%%%%%%%%%%%%%%%%%%%%%%%%%%%

\section{HVLB\_CC Algorithm}\label{sec:algorithm2}
Here, we describe the proposed HVLB\_CC static list scheduling algorithm in detail, including the task prioritizing, processor selection, and scheduling phases.

As described in Section \ref{sec:related_work2}, concurrent communication is allowed and network heterogeneity is not considered in most existing list scheduling algorithms. Moreover, task and message scheduling is not synchronized, which can influence schedule length. Thus, such scheduling schemes are unrealistic scheduling for automotive embedded systems. As mentioned previously, HSV\_CC \cite{hsv_cc} has been proposed to address these issues. However, as described in Section \ref{sec:problem2}, HSV\_CC cannot be applied directly to automotive DSMSs due to inefficient resource usage, the possibility that SPGs cannot be scheduled, and its inability to handle stream processing if the execution times varies. The proposed HVLB\_CC algorithm addresses these issues and can be applied to automotive DSMSs. 

In the task prioritizing phase, tasks in an SPG are prioritized according to HVLB\_CC values to prevent a high-depth task from being scheduled earlier than its predecessor at the lower depth. In the processor selection phase, tasks are allocated to appropriate processors while considering processor load. Tasks and messages are scheduled to the corresponding processor and communication links with specific start and finish times.

\subsection{Task Prioritizing Phase}
In the task prioritizing phase, a {\it rank} value must be calculated and assigned to each node. The rank value is calculated recursively by traversing the task graph (i.e., the DAG), starting from the exit node. Thus, it is sometimes referred to as the {\it upward rank}; however, we simply refer to it as the {\it rank}. The computation of the {\it rank} value in most existing list scheduling algorithms uses the average computation and communication times. However, calculating the {\it rank} value using average times cannot produce accurate scheduling results in heterogeneous computing and networking environments \cite{hsv_cc}. Therefore, we employ different computation and communication times for the different heterogeneous processors and communication links in the proposed method to calculate the {\it rank} value more accurately, which can be expressed as follows:
\begin{equation}
\label{eq:rank}
\hspace*{-3mm}
\begin{cases}
\; {\it rank}(n_i,p_{src}) \\\quad= {\it comp}(n_i,p_{src})  + \max\limits_{n_j\in succ(n_i)}\{{\it rank}(n_j,p_{src})+comm_{i,j}^{p_{src}} \}; \\
\; {\it rank}(n_{exit},p_{src}) = {\it comp}(n_{exit},p_{src});
\end{cases} 
\end{equation}
where ${\it comm}_{i,j}^{p_{src}}$ denotes the communication time from $n_i$ located on source processor $p_{src}$ to $n_j$. A higher {\it rank} indicates that the corresponding node has higher criticality and more influence on the execution of other nodes; thus, a node with a higher {\it rank} will be scheduled earlier than a node with lower {\it rank}.

To calculate the communication speed of source processor $p_{src}$, we must first compute the route $speed(R^{p_{src},p_{dest}})$ between the source processor $p_{src}$ and the destination processor $p_{dest}$, which can be expressed as follows:

\begin{equation}
\label{eq:speed}
\hspace*{-7mm}  {\it speed}(R^{p_{src},p_{dest}}) = \frac{1}{z}\mysum_{t=1}^{z} {\it speed}(r_t^{p_{src},p_{dest}}).
\end{equation}
Here, $z$ sub-routes exist between source processor $p_{src}$ and destination processor $p_{dest}$. A route between $p_{src}$ and $p_{dest}$ consists of one or more sub-routes and can be expressed as $R^{p_{src},p_{dest}}=\{r_1^{p_{src},p_{dest}},r_2^{p_{src},p_{dest}},...,r_z^{p_{src},p_{dest}}\}$. Equation \ref{eq:speed} indicates the average values of the speeds for all sub-routes between $p_{src}$ and $p_{dest}$, i.e., the speeds of the communication routes between $p_{src}$ and $p_{dest}$.

The speed of a route consisting one or more sub-routes can be determined using the minimum speed communication link because the message transmission time through multiple links is dependent on the link with the lowest speed. Thus, we can define communication route speed between $p_{src}$ and $p_{dest}$ as the average of the lowest speed value for all of the sub-routes in the communication route. This communication route speed can be expressed as follows:

\begin{equation}
\label{eq:speed_2}
\hspace*{-7mm} {\it speed}(R^{p_{src},p_{dest}}) = \frac{1}{z}\mysum_{t=1}^{z} \mymin_{l_v\in L}\{{\it speed}(l_v^t)\}.
\end{equation}

\begin{table*}[t]
\begin{minipage}{1.0\hsize}
\centering
\caption{{\it rank}, {\it hrank}, {\it outd}, {\it depth}, and {\it HPRV} values (Fig. \ref{fig:dsg_sample})}
\label{tab:values_rank}\hspace*{-7mm}
\scriptsize
\begin{tabular}{p{2.5cm}p{1.0cm}p{1.0cm}p{1.0cm}p{1.0cm}p{1.0cm}p{1.0cm}p{1.0cm}p{1.0cm}p{1.0cm}p{1.0cm}} \hline
Task&$n_1$&$n_2$&$n_3$&$n_4$&$n_5$&$n_6$&$n_7$&$n_8$&$n_9$&$n_{10}$\\[1.0ex] \hline
$rank(n_i,p_1)$&145.0 &133.0 &109.0 &109.0 &85.0 &50.0 &67.0 &48.0 &20.0 &15.0\\
$rank(n_i,p_2)$&81.66 &74.99 &61.66 &61.66 &48.33 &29.67 &38.33 &28.0 &13.0 &10.0\\
$rank(n_i,p_3)$&96.99 &90.33 &73.67 &73.67 &57.0 &36.0 &45.33 &34.33 &16.0 &12.0\\
$hrank(n_i)$&107.9 &99.4 &81.4 &81.4 &63.4 &38.6 &50.2 &36.8 &16.3 &12.3 \\
$outd(n_i)$&2 &2 &2 &2 &2 &1 &1 &1 &0 &0 \\
${\it depth}(n_i)$&1 &1 &1 &2 &2 &2 &3 &3 &4 &4 \\
${\it HPRV}(n_i)$&215.8 &198.8 &162.8 &162.8 &126.8 &38.6 &50.2 &36.8 &0 &0\\
${\it HPRV}(n_i,G)$&107.9 &99.4 &81.4 &20.4 &15.9 &9.7 &5.6 &4.1 &0 &0\\
\hline\end{tabular}
\end{minipage} 
\end{table*}

\begin{table}[t]
\begin{center}
\centering
\caption{Example of communication speeds of routes}
\label{tab:comm_speed}
\footnotesize
%\scriptsize
\begin{tabular}{p{1.1cm}cccccc} \hline
Route&$R^{p_1,p_2}$&$R^{p_1,p_3}$&$R^{p_2,p_3}$&$R^{p_2,p_1}$&$R^{p_3,p_1}$&$R^{p_3,p_2}$\\[0.7ex] \hline
Speed&1&1&2&1&1&2\\
\hline\end{tabular}
\end{center}
\vspace{-0.3cm}
\end{table}

Assume that there are two sub-routes between $p_2$ and $p_3$, $r_1^{p_2,p_3}=\{l_2,l_4\}$ and $r_2^{p_2,p_3}=\{l_3\}$. The speeds of the sub-routes are 1 and 3 for $r_1^{p_2,p_3}$ and $r_2^{p_2,p_3}$, respectively, because the minimum link speed in the sub-route is the communication route speed. The speeds for the other routes can be calculated in the same way and are shown in \textbf{Table} \ref{tab:comm_speed}.

Then, we compute the data transfer speed of the source processor, ${\it speed}(p_{src})$, which is the average value of all of the communication route speeds between the source processor and one or more destination processors and can be expressed as follows: 

\begin{equation}
\label{eq:speed_3}
 \hspace*{-7mm} {\it speed}(p_{src}) = \frac{1}{|P|-1}\mysum_{p_{dest}\in P,p_{dest}\neq p_{src}} speed(R^{p_{src},p_{dest}}).
\end{equation}

Note that the data transfer speeds of other source processors can be calculated using Eq. \ref{eq:speed_3}. For $p_1$, $p_2$, and $p_3$, the speeds are 1.0, 1.5, and 1.5, respectively.

The communication time, ${\it comm}_{i,j}^{p_{src}}$, required to transmit data from task $n_i$ on $p_{src}$ to task $n_j$ can be computed by dividing the message size by the data transfer speed of the source processor. It can be expressed as follows:

\begin{equation}
\label{eq:comm_time}
\hspace*{-7mm} {\it comm}_{i,j}^{p_{src}} = \frac{{\it tpl}(e_{i,j})}{{\it speed}(p_{src})},
\end{equation}
where $speed(p_{src})$ is the data transfer speed of the source processor defined by Eq. \ref{eq:speed_3}. If $n_i$ and $n_j$ are on the same processor, then the communication time between them is considered to be negligible. 

In existing algorithms \cite{list_advantage,clus1,clus2,clus3,clus4,clus5}, the average processor speed is used because these algorithms assume that communication contention has not occurred; thus, the same communication times are used for all heterogeneous processors. In contrast, we consider communication contention by defining the processor data transfer speed, while most existing algorithms use the same data transfer times for of the heterogeneous processors. This principle is based on HSV\_CC. 

In addition, {\it rank} has different values for different heterogeneous processors, as shown in Eq. \ref{eq:rank}. The average {\it rank} values for all possibly assign processors represented by {\it hrank} is defined as follows:

\begin{comment}
\begin{table}[t]
\begin{center}
\centering
\caption{Example of communication speeds of processors}
\label{tab:comm_speed_proc}
\footnotesize
\begin{tabular}{p{1.2cm}ccc} \hline
Processor&$p_1$&$p_2$&$p_3$\\[0.7ex] \hline
Speed&1.0&1.5&1.5\\
\hline\end{tabular}
\end{center}
\vspace{-0.3cm}
\end{table}
\end{comment}

\begin{equation}
\label{eq:hrank}
 \hspace*{-7mm} {\it hrank}(n_i) = \frac{1}{|P|} \times \mysum_{p_u\in P}rank(n_i,p_u).
\end{equation}

While priority is given to each task, HSV\_CC considers {\it hrank} and {\it outd} because considering the out-degree of a node can minimize the schedule length simultaneously \cite{hsv_cc}. The heterogeneous priority rank value (i.e., $HPRV\_{CC}$) used for task prioritization is defined as 

\begin{equation}
\label{eq:hprv1}
\hspace*{-7mm}{\it HPRV\_CC}(n_i) = hrank(n_i) \times outd(n_i).
\end{equation}

{\it HPRV\_CC} is used in the proposed algorithm, which we refer to as HVLB\_CC (A). In the {\it HPRV\_CC} equation, a node with many immediate successor nodes can have a higher probability of being assigned earlier. However, Eq. \ref{eq:hprv1} can restrict scheduling for various task graphs because only task graphs with $HPRV\_CC(n_p)\geq HPRV\_CC(n_s)$ are used, where $n_p$ and $n_s$ are predecessor and successor tasks, respectively. If the {\it hrank} values of these two tasks do not change significantly and $n_s$ has a large number of out-degrees, then $HPRV\_CC(n_p)<HPRV\_CC(n_s)$ can occur. 

In stream processing, successor tasks located at high depths in the task graph can have greater $outd$ values than those of their predecessor tasks located at lower depths. To prevent a high-depth task from being scheduled earlier than its predecessor at a lower depth, we define HPRV\_CC (B) as follows:  

\begin{equation}
\begin{split}
\label{eq:hprv2}
\hspace*{-7mm}{\it HPRV\_CC}(&n_i,G) 
 \\&= hrank(n_i) \times \frac{outd(n_i)}{\max\{outd(G)\}} \times \frac{1}{depth(n_i)^2},
\end{split}
\end{equation}
where $\max\{outd(G)\}$ indicates the maximum out-degree value of the task graph $G$. As shown in Eq. \ref{eq:hprv2}, we decrease the influence of $outd$ on task prioritizing and consider the {\it depth} of each task in the task graph to prevent assigning a higher priority to a successor task located at a higher depth\footnote{We explain why we use $depth(n_i)^2$ in Eq. \ref{eq:hprv2} in Section 5.}. The proposed algorithm that employs Eq. \ref{eq:hprv2} is referred to as HVLB\_CC (B).

\subsection{Processor Selection Phase}
In this phase, the ordered tasks in the priority queue are assigned to appropriate processors that can provide them with the smallest possible {\it HVLB\_CC} values. In HSV\_CC, {\it EFT} and the longest distance exit time with communication contention ({\it LDET\_CC}) are important factors in processor selection. However, these two factors are highly dependent on the capabilities of processors and communication links. Inefficient schedule decisions in which tasks and messages tend to be assigned to high-capability processors and links may occur (Section \ref{sec:problem2}). As a result, HSV\_CC cannot prevent specific processors and links from becoming congested with tasks and messages. Thus, we propose an approach that considers processor load while allocating tasks to appropriate processors so that tasks and messages are not concentrated on specific resources. By considering both of the factors used in HSV\_CC and the processor loads, the proposed algorithm can decrease the schedule length and balance the processor load better than HSV\_CC can.

The earliest start time ({\it EST}) of a task is determined by the processor capability (i.e., the speed) and the communication route from the source processor to the destination processor. For an entry node, the value of {\it EST} for a node will be 0 if the entry task is scheduled on the processor at the first time. If any node has already been scheduled, {\it EST} equal its maximum value of 0 and available processor time. Therefore, {\it EST} is defined by Eq. \ref{eq:est}. With the exception of the entry nodes, each node must consider the time at which message transmission from its predecessor is completed:

\begin{equation}
\label{eq:est}
\hspace*{-7mm}
\begin{cases}
\; {\it EST}(n_{entry},p_{dest},r_{z}^{p_{src},p_{dest}}) = \max\{0,avail[p_{dest}]\};\\
\; {\it EST}(n_{j},p_{dest},r_{z}^{p_{src},p_{dest}}) = {\color{black}\max}\{avail[p_{dest}],\\ \mymax_{n_i\in pred(n_j),proc(n_i)=p_{src},p_{src}\in P}\{{\it MFT}(e_{i,j}^{p_{src},p_{dest}}, r_z^{p_{src},p_{dest}})\} \};
\end{cases} 
\end{equation}

where ${\it MFT}(e_{i,j}^{p_{src},p_{dest}}, r_z^{p_{src},p_{dest}})$ is the time at which the message transmission on sub-route $r_z^{p_{src},p_{dest}}$ is completed, and { \color{black}$avail[p_{dest}]$ is the available time of processor $p_{dest}$}. {\it MFT} can be determined using the finish time of the last link $l_{end}^z$ of sub-route $r_z^{p_{src},p_{dest}})$. Therefore, {\it EST} can be calculated using the link finish time ({\it LFT}) rather than {\it MFT}, as follows:

\begin{equation}
\label{eq:est_2}
\hspace*{-7mm}
\begin{cases}
\; {\it EST}(n_{entry},p_{dest},r_{z}^{p_{src},p_{dest}}) = \max\{0,avail[p_{dest}]\};\\
\; {\it EST}(n_{j},p_{dest},r_{z}^{p_{src},p_{dest}}) = {\color{black}\max}\{avail[p_{dest}],\\ \mymax_{n_i\in pred(n_j),proc(n_i)=p_{src},p_{src}\in P}\{{\it LFT}(e_{i,j}^{p_{src},p_{dest}}, l_{end}^z, r_z^{p_{src},p_{dest}})\} \};
\end{cases} 
\end{equation}

{\it EFT} can be defined as follows for a task in a communication contention environment:

\begin{equation}
\begin{split}
\label{eq:eft}
\hspace*{-7mm}
{\it EFT}(n_{j},&p_{dest},r_{z}^{p_{src},p_{dest}}) \\&= {\it EST}(n_{j},p_{dest},r_{z}^{p_{src},p_{dest}}) + {\it comp}(n_j,p_{dest}).
\end{split}
\end{equation}

According to the above timing analysis, we can synchronize and schedule tasks and messages simultaneously. The value of {\it EST} for a task depends on the message transmission time, which in turn depends on the actual value of {\it LFT}. To obtain {\it LFT}, we must first compute the link start time ({\it LST}).

{\it LST}, the time at which message $e_{i,j}^{p_{src},p_{dest}}$ begins to be transmitted through sub-route $r_z^{p_{src},p_{dest}}$, can be expressed as follows:

\begin{equation}
\label{eq:lst}
\hspace*{-7mm}
\begin{cases}
\; {\it LST}(e_{i,j}^{p_{src},p_{dest}},l_1^z, r_z^{p_{src},p_{dest}}) = \max\{{\it AFT}(n_i), {\it avail}(l_1^z)\};\\
\; {\it LST}(e_{i,j}^{p_{src},p_{dest}},l_{x+1}^z, r_z^{p_{src},p_{dest}})\\ \qquad\quad = \max\{{\it LST}(e_{i,j}^{p_{src},p_{dest}},l_x^z,r_z^{p_{src},p_{dest}}), {\it avail}(l_{x+1}^z)\};
\end{cases} 
\end{equation}
%{\color{black}avail$(l_z^x)$ は $l_z^x$ が使用可能になる時間を表しています。}
where $AFT(n_i)$ is the actual finish time of task $n_i$, and { \color{black}$avail(l_z^x)$ is the available time of link $l_z^x$. } When {\it LST} is calculated using Eq. \ref{eq:lst}, we can obtain {\it LFT} as follows:

\begin{equation}
\label{lft}
\hspace*{-7mm}
\begin{cases}
\; {\it LFT}(e_{i,j}^{p_{src},p_{dest}},l_1^z, r_z^{p_{src},p_{dest}}) \\
\quad= {\it LST}(e_{i,j}^{p_{src},p_{dest}},l_1^z, r_z^{p_{src},p_{dest}}) + {\it CTML}(e_{i,j}^{p_{src},p_{dest}},l_1^z);\\
\; {\it LFT}(e_{i,j}^{p_{src},p_{dest}},l_{x+1}^z, r_z^{p_{src},p_{dest}}) \\
\quad= \max\{{\it LFT}(e_{i,j}^{p_{src},p_{dest}},l_x^z, r_z^{p_{src},p_{dest}}), \\
\; \quad{\it LST}(e_{i,j}^{p_{src},p_{dest}},l_{x+1}^z, r_z^{p_{src},p_{dest}}) + {\it CTML}(e_{i,j}^{p_{src},p_{dest}},l_{x+1}^z)\};\\
\end{cases} 
\end{equation}

where ${\it CTML}(e_{i,j}^{p_{src},p_{dest}},l_{x+1}^z)$ indicates the communication time to transmit message $e_{i,j}^{p_{src},p_{dest}}$ through link $l_{x+1}^z$. This time can be expressed as follows:

\begin{equation}
\label{eq:ctml}\hspace*{-7mm}
{\it CTML}(e_{i,j}^{p_{src},p_{dest}},l_x^z) = \frac{tpl(e_{i,j}^{p_{src},p_{dest}})}{{\it speed}(l_x^z)}
\end{equation}

where ${\it speed}(l_x^z)$ represents link $l_x^z$'s communication speed and 
$tpl(e_{i,j}^{p_{src},p_{dest}})$ is the amount of a message transmitted from the source processor to the destination processor.

In HSV\_CC, the {\it downward} factor (i.e., {\it EFT}) and the {\it upward} factor (Eq. \ref{eq:ldet}) are considered:

\begin{equation}
\label{eq:ldet}\hspace*{-7mm}
LDET\_CC(n_i,p_u) = rank(n_i,p_u) - {\it comp}(n_i,p_u).
\end{equation}

By considering both factors in the processor selection criteria, HSV\_CC achieves overall schedule lengths that are superior to those of existing algorithm \cite{hsv_cc}. Note that LDET\_CC will be 1.0 for exit nodes.

\theoremstyle{definition}
\begin{definition}{Balancing Parameter (BP)}\label{def:bp}
\begin{equation}
\label{eq:bp}\hspace*{-7mm}
BP(p_{dest},\alpha) = 1.0 + \left(\frac{\mysum_{k=1}^mcomp(n_k,p_{dest})}{period}\right) \times \alpha
\end{equation}
\end{definition}

Here, $m$ is the number of tasks assigned to the destination processor $p_{dest}$, and the numerator indicates the cumulative processing time of $m$ tasks. $period$ represents the DAG (application) period and deadline. $BP$ begins at 1.0, and the CPU utilization of each processor is calculated based on the cumulative computation times of the assigned tasks to $p_{dest}$. The parameter $\alpha$ is used to adjust the amount of $BP$ influence on the value of {\it HVLB\_CC}. 

\theoremstyle{definition}
\begin{definition}{Heterogeneous Value with Load Balancing and Communication Contention}\label{def:hvlb}
\begin{equation}
\label{eq:hvlb}\hspace*{-2mm}
\begin{cases}
\; HVLB\_CC(n_j,p_{dest},r_z^{p_{src},p_{dest}}) =\\
EFT(n_j,p_{dest},r_z^{p_{src},p_{dest}}) \times LDET\_CC(n_i,p_{dest}) \times BP(p_{dest},\alpha);\\
\;HVLB\_CC(n_{exit},p_{dest},r_z^{p_{src},p_{dest}}) = EFT(n_{exit},p_{dest},r_z^{p_{src},p_{dest}}); 
\end{cases} 
\end{equation}
\end{definition}

The proposed algorithm can consider {\it downward} and {\it upward} simultaneously and can balance loads over multiple processors. This is the most significant improvement over HSV\_CC. {\it BP} helps balance the processor load by giving low-load processors the chance to be selected. For a task scheduled at last, we consider only the {\it EFT} value to reduce the overall schedule length. 

\subsection{Scheduling Algorithm}
We present a heuristic task and message scheduling approach (i.e., HVLB\_CC) to solve the unbalanced assignment problem that occurs in HSV\_CC. To consider LB while scheduling dependent tasks over multiple processors, the HVLB\_CC algorithm searches for the optimal task and message assignments by adjusting $\alpha$ which is described in \textbf{Definition} \ref{def:bp}. 

\begin{figure}
\centering
\includegraphics[width=1.0\linewidth]{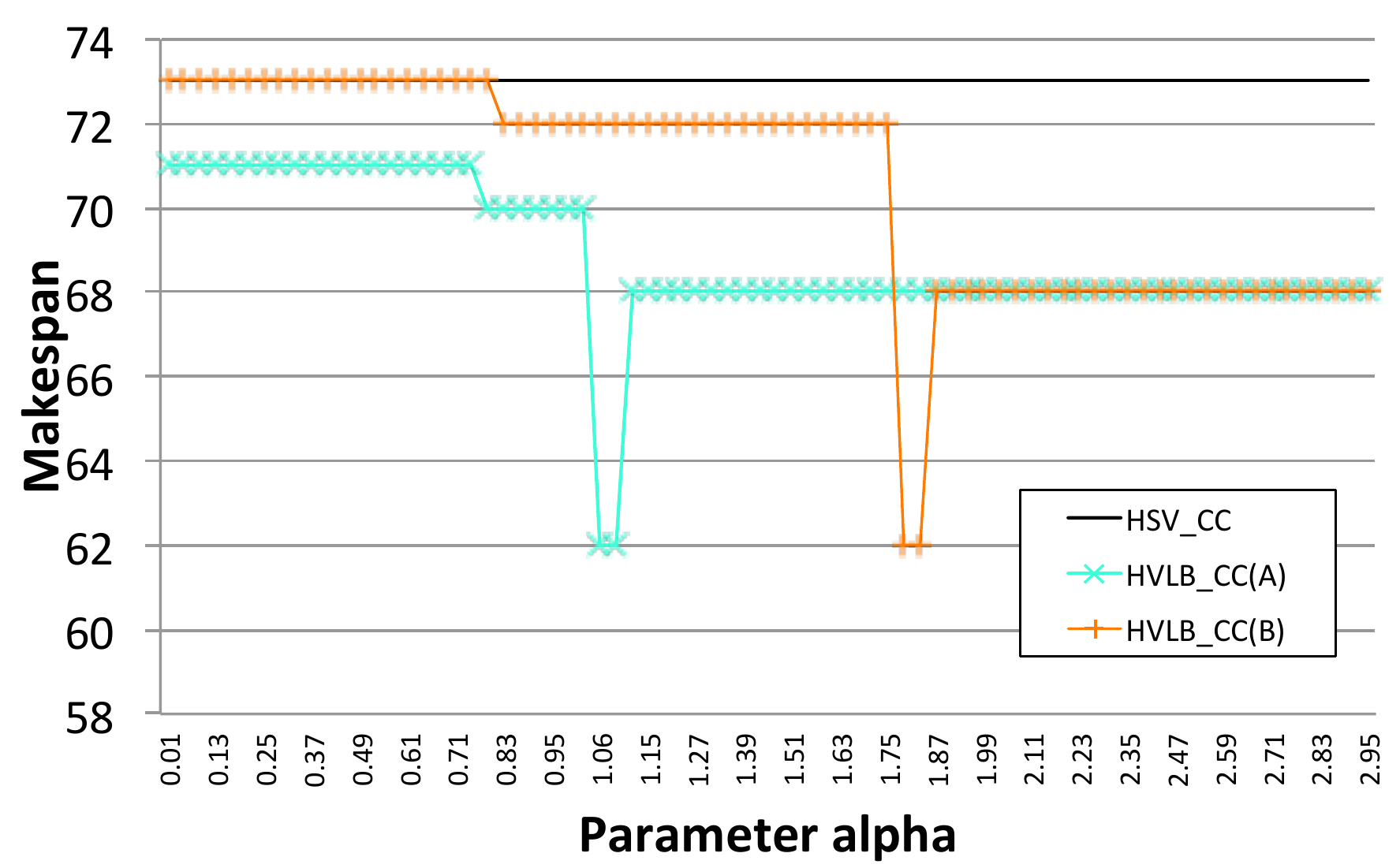}
\vspace{-0.1cm}

\caption{\color{black}Effect of parameter $\alpha$ on schedule length in proposed algorithms}
\label{fig:example_makespan}
\vspace{-0.3cm}
\end{figure}

As shown in \textbf{Figure} \ref{fig:example_makespan}, we can search for the optimal assignment that produces a value of {\it makespan} smaller than the value that can be produced by HSV\_CC. We initiated $\alpha$ at zero and increased its value from 0.01 to 3.0. In addition, the {\it CCR} was fixed to 1 in this example. For HVLB\_CC (A), when $\alpha$ is in the range [0.0, 0.79], {\it makespan} is 73, which is the same as its value for HSV\_CC. {\it makespan} is 72 in the range [0.8, 1.76], and we can obtain the minimum {\it makespan} value of 62 in the range [1.77, 1.83]. From 1.84 to 3.0, {\it makespan} is constant at 68. Similarly, the minimum values of {\it makespan} are 71, 70, 62, and 68 in the ranges [0.0, 0.78], [0.79, 1.05], [1.06, 1.1], and [1.11, 3.0], respectively, for HVLB\_CC (B). Then, the proposed algorithms selected the scheduling results that could produce the minimum {\it makespan} value with $\alpha$ in the ranges [1.77, 1.83] and [1.06, 1.1] for HVLB\_CC (A) and HVLB\_CC (B), respectively. 

\begin{algorithm}[t]
\KwIn{Application (SPG), the number of processors, and network topology}
\KwOut{Assignment result for the SPG, i.e., start times and finish times are determined for all tasks and messages}
Calculate {\it rank}, {\it hrank}, {\it LDET\_CC}, and {\it HPRV\_CC} for each task\;
Calculate {\it depth} for each task; // if HVLB\_CC (B) is used \\
Enqueue tasks into a priority queue according to a nonincreasing order of {\it HPRV\_CC}\;
$\alpha$ = 0\;
\For{$cycle \gets 1$ \textbf{to} $k$} {
\While{not all tasks in the priority queue are scheduled} {
  Dequeue the task $n_i$ has the maximum {\it HPRV\_CC}\;
  Compute {\it BP} using Eq. \ref{eq:bp} in all processors\;
  Compute the {\it HVLB\_CC} using Eq. \ref{eq:hvlb} for $n_i$ in all processors\;
  Schedule $n_i$ on to the corresponding processor and the messages to the corresponding links that minimize {\it HVLB\_CC}\;
  Mark $n_i$ as a scheduled task\;
}
Compute the {\it makespan} of the assignment result\;
\If{the {\it makespan} is the minimum} {
	$minimum\ alpha \gets \alpha$\;
}
$\alpha \gets \alpha + 0.01$\;
}
\caption{HVLB\_CC Algorithm}
\label{alg:hvlb}
\end{algorithm}

\begin{figure}
\centering
\includegraphics[width=1.01\linewidth]{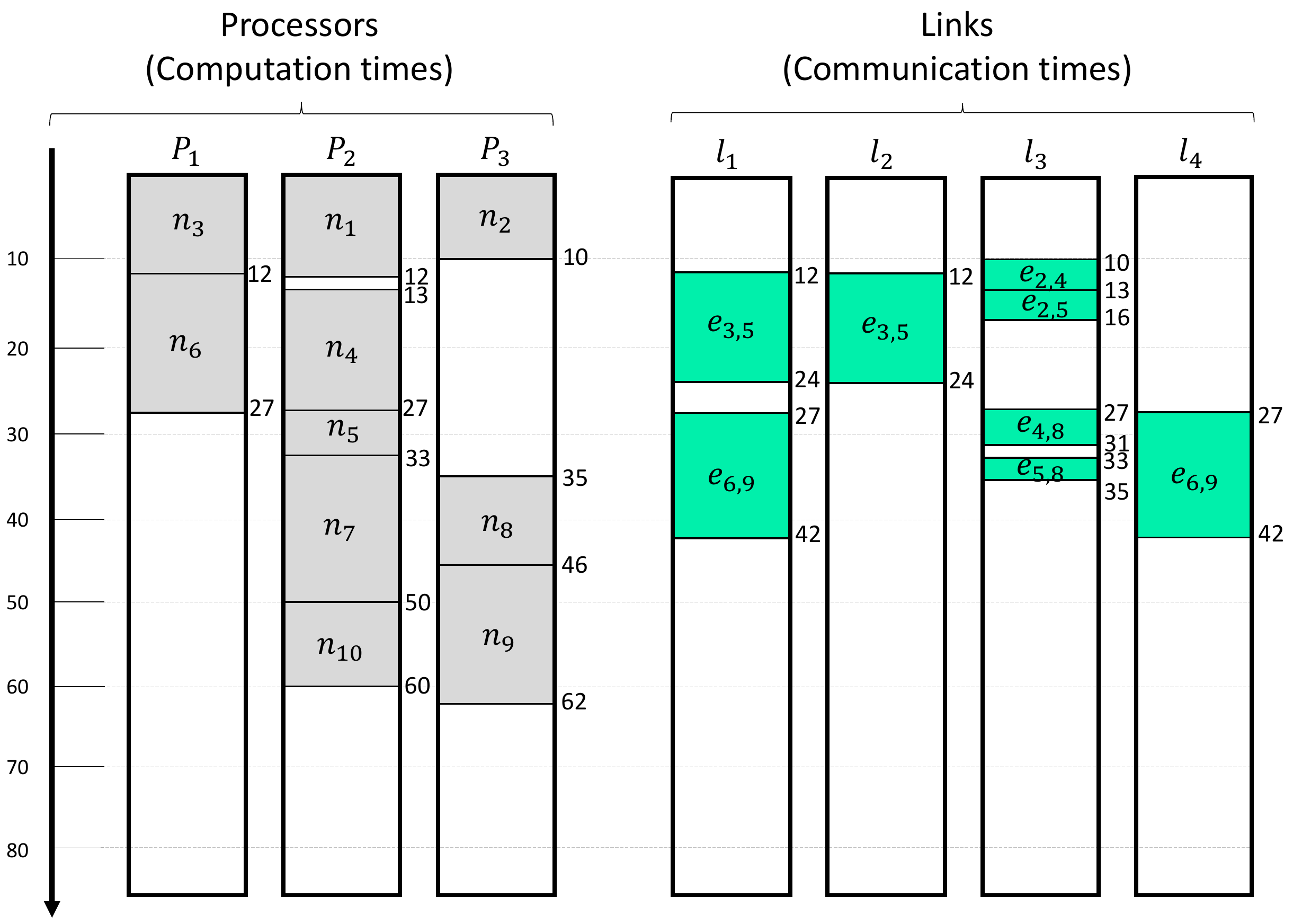}
\caption{Gantt chart of SPG scheduling in Fig. \ref{fig:dsg_sample} with the proposed HVLB\_CC ({\it makespan}=62)}
\label{fig:hvlb_cc_assign}
\vspace{-0.3cm}
\end{figure}

The proposed HVLB\_CC approach is described in \textbf{Algorithm} \ref{alg:hvlb}. The task order in the priority queue in HVLB\_CC (A) is $\{n_1,n_2,n_3,n_4,n_5,n_7,n_6,n_8,n_9,n_{10}\}$, which is the same as it is in HSV\_CC. In HVLB\_CC (B), the task order in the priority queue is $\{n_1,n_2,n_3,n_4,n_5,n_6,n_7,n_8,n_9,n_{10}\}$. The $\alpha$ values that can record the minimum {\it makespan} values are 1.77 and 1.06 in HVLB\_CC (A) and HVLB\_CC (B), respectively. The minimum values of {\it makespan} in scheduling results obtained using HVLB\_CC (A) and HVLB\_CC (B) are both 62, while {\it makespan} is 73 for HSV\_CC. Thus, the proposed HVLB\_CC algorithm yields smaller values of {\it makespan} and improves the load balance between three processors in comparison to the HSV\_CC algorithm. A Gantt chart of the scheduling results obtained using the proposed HVLB\_CC algorithm is shown in \textbf{Figure} \ref{fig:hvlb_cc_assign}. Note that the scheduling results at the minimum {\it makespan} values for HVLB\_CC (A) and HVLB\_CC (B) are the same.

Here, we discuss the time complexity of the proposed heuristic HVLB\_CC algorithm, which can be evaluated as follows. Assume that there are $n$ computation tasks and $p$ networked processors. {\it rank}, {\it hrank}, and {\it HPRV\_CC} must be calculated for all of the tasks, and these values must be compared for all of the processors, which can be performed within $O(p\times n)$. Note that scheduling must traverse all of the tasks and can be performed within $O(n)$. The scheduling to find the minimum {\it makespan} value in the range of $\alpha$ can be performed within $O(k)$. Therefore, the complexity of the proposed HVLB\_CC algorithm is $O(p\times n^2\times k)$ whereas that of HSV\_CC is $O(p\times n^2)$ \cite{hsv_cc}. However, the proposed algorithm can produce more accurate schedule results and decrease {\it makespan}.

{ \color{black}In the time complexity of the proposed algorithm, parameter $k$ increases in the range of $\alpha$, which is used to adjust the value of $BP$ (the balancing parameter) relative to that of HVLB\_CC. The parameter $k$ in $O(p \times n^2 \times k)$ will not affect the entire time complexity significantly compared to the value of parameter $n^2$ because $n^2$ requires greater computational time. Even though the time complexity of the proposed algorithm is greater than that of HSV\_CC, our algorithm can provide more accurate scheduling results and decreased makespan.}

\if0
\begin{algorithm}[h]
\KwIn{Application (SPG), the number of processors, and network topology}
\KwOut{Assignment result for the SPG, i.e., start times and finish times are determined for all tasks and messages}
Calculate {\it rank}, {\it hrank}, {\it LDET\_CC}, and {\it HPRV\_CC} for each task\;
Calculate {\it depth} for each task; // if HVLB\_CC(B) is used \\
Enqueue tasks into a priority queue according to a nonincreasing order of {\it HPRV\_CC}\;
$\alpha$ = 0\;
\For{$cycle \gets 1$ \textbf{to} $k$} {
\While{not all tasks in the priority queue are scheduled} {
  Dequeue { \color{black}the task $n_i$ has the maximum} {\it HPRV\_CC}\;
  Compute {\it BP} using Eq. \ref{eq:bp} in all processors\;
  Compute the {\it HVLB\_CC} using Eq. \ref{eq:hvlb} for $n_i$ in all processors\;
  Schedule $n_i$ on to the corresponding processor and the messages to the corresponding links that minimize {\it HVLB\_CC}\;
  Mark $n_i$ as a scheduled task\;
}
Compute the {\it makespan} of the assignment result\;
\If{the {\it makespan} is the minimum} {
	$minimum\ alpha \gets \alpha$\;
}
$\alpha \gets \alpha + 0.01$\;
}
\caption{HVLB\_CC Algorithm}
\label{alg:hvlb}
\end{algorithm}

\fi

\subsection{Imprecise Computation Model}
As mentioned earlier, the proposed HVLB\_CC\_IC scheduling algorithm utilizes imprecise computations. We assume that a premature task can record its latest intermediate result when it terminates execution. Note that not all of the tasks in an SPG are applied in the imprecise computations model; instead, only those tasks whose input data rates vary are employed. { \color{black}An example of a task for which the amount of required processing fluctuates is a task that requires V2V communication to exchange data that changes in amount according to the surrounding situation.
} The computation time of each task adopted in the imprecise computation model is assumed to consist of a {\it mandatory part} $mp_i$, which is followed by an {\it optional part} $op_i$. This can be defined as follows:

\begin{equation}
\label{eq:impre_comp}\hspace*{-7mm}
  comp(n_{i},p_u) = mp_i+op_i
\end{equation}
where $0<mp_i<comp_i$. { \color{black}For example, the process required to determine the navigation in the navigation system is the mandatory part of the model and the road-to-vehicle communication is the optional part.}
An execution of the {\it optional part} can begin after the processing {\it mandatory part} has been processed. In case an imprecise computation model task finishes executing only its {\it mandatory part} or terminates before completing execution of the {\it optional part}, then the results of the task are {\it imprecise}. In contrast, the results of a task are {\it precise} when the task processes all of the {\it mandatory} and {\it optional} parts. 

\subsubsection{Searching for schedule holes}
The schedule results produced by the proposed algorithm can have possible idle time slots called schedule holes. Due to the precedence constraints between tasks, schedule holes may occur in the schedule of a particular processor. For example, for a task that receives two pieces of input data from each predecessor task, the times at which the pieces of data are generated by the two predecessor tasks are generally not the same. Thus, if the data arrives from one predecessor task much earlier than it does from the other task, the processor idle time slots can occur between the time at which the data is sent from the predecessor task and the time at which all of the required data arrive at the successor task. Note that not all of the processor idle time slots are available for execution of an {\it optional part}.

Schedule holes that can be utilized for an {\it optional part} of a task can be calculated using the following equations. If a processor with an allocated predecessor node $n_p$ is the same as the processor with successor node $n_s$, the schedule hole between them can be calculated as follows:

\begin{equation}
\label{eq:condition1}\hspace*{-7mm}
\begin{split}
%\begin{align}
{\it condition1}(n_p)=\min_{n_s\in succ(n_p),proc(n_p)=proc(n_s)}\{\min\{{\it EST}(n_s),\\{\it EST}(n_{np})\}-\left({\it EST}(n_p)+{\it comp}(n_p,p_{src})\right)\}>0.
%\end{align}
\end{split}
\end{equation}
where $n_{np}$ is a node executed after $n_p$ on the $proc(n_p)$.

If a processor with an allocated precedence node differs from the processor with the successor node, the schedule hole between them can be calculated as follows:

\begin{equation}
\label{eq:condition2}\hspace*{-7mm}
\begin{split}
&{\it condition2}(n_p)=\min_{n_s\in succ(n_p),proc(n_p)\neq proc(n_s)}\{\min\{{\it EST}(n_s),\\&{\it EST}(n_{np}),{\it LST}''(e_{p,s}^{p_{src}})\}-({\it EST}(n_p)+{\it comp}(n_p,p_{src}))\}>0,
\end{split}
\end{equation}
where ${\it LST}''(e_{p,s}^{p_{src}})$ is the recalculated value of {\it LST} that is delayed to time that can be maximized the schedule hole and must not influence the data receive time to successor node.  

A task with one successor task allocated to the same processor or a different processor with its own task must satisfy either {\it condition 1} or {\it condition 2}. For a task with more than two successor tasks, schedule holes occurs for the execution of the {\it optional part} when both {\it condition 1} and {\it condition 2} are satisfied. The maximum amount of available processor time slots will be the minimum value of the results of {\it condition 1} and {\it condition 2}.

\section{Performance Evaluation}\label{sec:simulation2}
The performance evaluation was conducted in five parts. We compared the existing algorithm (HSV\_CC) and the proposed algorithms (HVLB\_CC (A) and HVLB\_CC (B)) in terms of (i) the schedule length ratio ({\it SLR}) and {\it speedup} with increasing number of tasks, (ii) LB, (iii) the value of {\it SLR} with increasing communication to computation ratio ({\it CCR}), (iv) the scheduling failure rate, and (v) the data precision in HVLB\_CC with and without an imprecise computation model.

\subsection{Experimental Metrics}
The performance comparisons of the algorithms were based on {\it SLR} and {\it speedup} which are used in \cite{hsv_cc}, and LB \cite{lb}. {\it SLR} is a normalized schedule length, and it is computed by dividing the schedule length ({\it makespan}) by the minimum execution time for all of the tasks in the critical path (CP) among multiple processors. It is defined as follows:
\begin{equation}
\label{eq:slr}\hspace*{-7mm}
  {\it SLR} = \frac{{\it makespan}}{\mysum_{n_i\in CP}\mymin_{p_u\in p}\left[{\it comp}_{(n_i,p_u)}\right]}.
\end{equation}

The {\it speedup} value for a given schedule is computed by dividing the minimum sequential execution time by the {\it makespan} value of the parallel execution time and is defined as follows:
\begin{equation}
\label{eq:speedup}\hspace*{-7mm}
  {\it speedup} = \frac{\mymin_{p_u\in p}\left[\mysum_{n_i\in N}{\it comp}({n_i,p_u})\right]}{{\it makespan}}
\end{equation}
where the minimum sequential execution time is computed by assigning all of the tasks to a single processor that can minimize the cumulative computation time.

LB is computed by dividing {\it makespan} by the ratio of the sum of processing time of each processor and the number of processors ({\it Avg}) and can be defined as follows:
\begin{equation}
\label{eq:loadbalancing}\hspace*{-7mm}
  LB = \frac{{\it makespan}}{{\it Avg}}
\end{equation}
where {\it Avg} is the average execution time over all processors. {\it Avg} can be calculated by using
\textcolor{black}{
\begin{equation}
\label{eq:loadbalancing}\hspace*{-7mm}
 {\it Avg} = \frac{\textstyle\sum\limits_{k=1}^{n}{\textstyle\sum\limits_{n_i\in N_k}}comp(n_i,p_k)}{n}
\end{equation}
}
where $n$ is the number of processors. { \color{black}$N_k$ represents a set of tasks allocated to the processor $p_k$}. Note that the numerator in this equation indicates the sum of computation times of $m$ allocated tasks for all $n$ processors.

%%%%%%%%%%%%%%%%%%%%%%%%%%%%%%%%%%%%%%%%%%%%%%%%%%%%%%%%%%%%%%%%%%
\subsection{Simulation Setup}
The task graphs for our experiments were generated by using Task Graphs For Free (TGFF) version 3.5 \cite{tgff}. TGFF can generate random DAGs for various parameters, such as number of tasks and maximum in-degree and out-degree. We set the experimental parameters as follows: number of tasks $n=\{10,20,30,40,50\}$, maximum in-degree $i=2$, and maximum out-degree $o=3$, and the minimum number of entry and exit nodes was 2. In experiments (i)-(iii), we generated random task graphs with the constraint $outd(n_p)\geq outd(n_s)$, where $n_s$ uses the results from $n_p$, because HSV\_CC fail to schedule task graphs without the out-degree constraint. These parameters were determined by considering the actual operator sets used in an automotive DSMS. 

{\it CCR} is the communication-to-computation ratio. A low value of {\it CCR} in a task graph can be considered as computation intensive processing (application). If $CCR > 1$ or higher, the communication time is greater than the computation time. In addition, the heterogeneity of communication becomes more obvious. On the other hand, the processing is communication-intensive if {\it CCR} is high. If $CCR < 1$, computation dominates the systems. In these experiments, we set $CCR=\{0.1, 0.5, 1.0, 5.0, 10.0\}$.

Three processors with different capabilities were used in all of the experiments. The execution rate $\mu$ of each processor was selected from $[0.67, 0.83, 1.0]$; thus, there were six execution rate patterns for the processors\footnote{To simplify the comparison, the experimental results of only three patterns of the processor execution rates are given in this paper.}. Note that the network topology\footnote{The network topology is the same as that one presented in \cite{hsv_cc} to compare the scheduling performance with that of the existing algorithm.} shown in \textbf{Figure} \ref{fig:network_topology} was applied in all of the experiments.

\begin{figure*}[p]
\begin{center}
\subfloat[processor execution rate of (1.0,0.67,0.83)]{%
\includegraphics[width=0.3\linewidth]{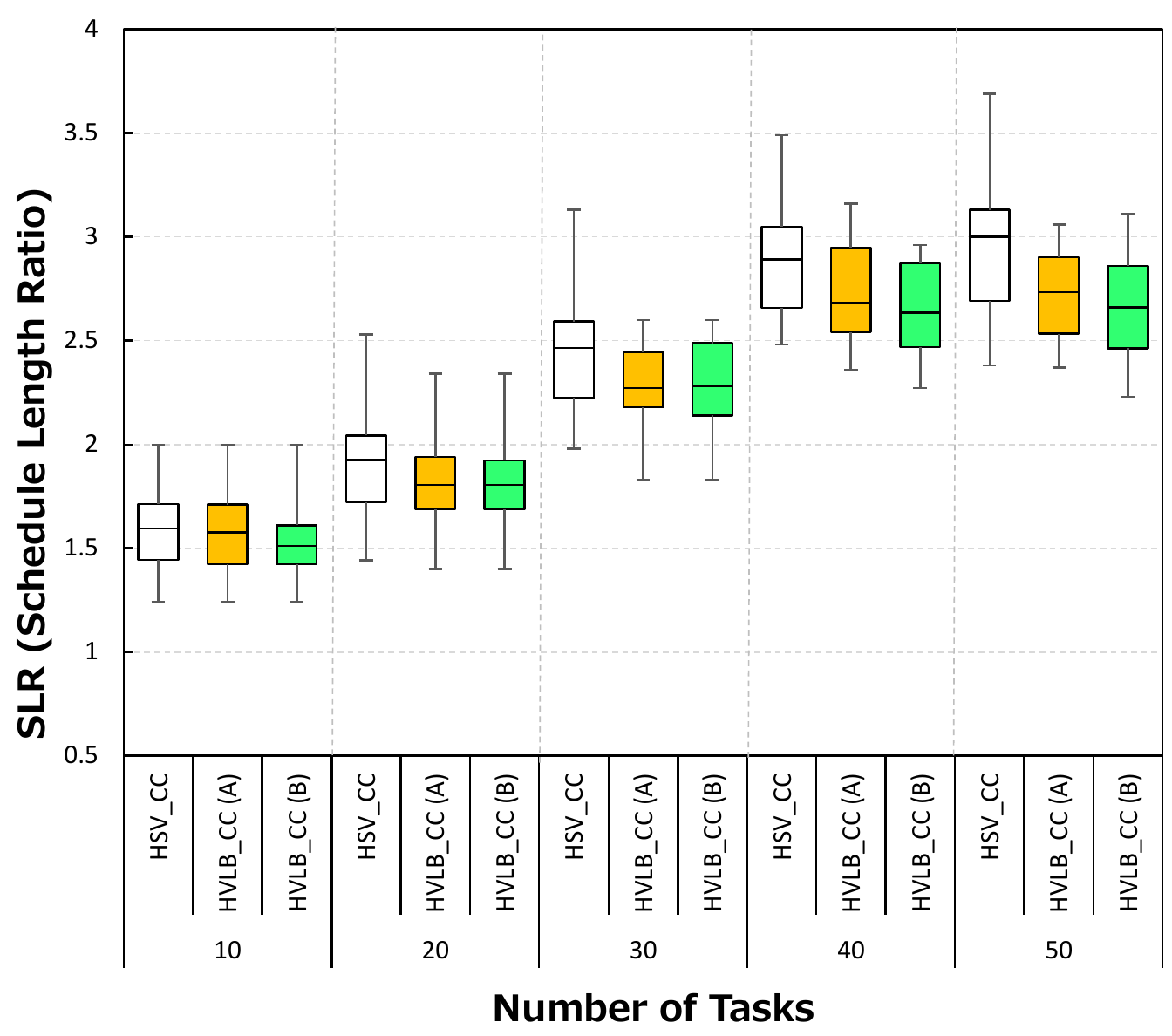}}%
\hspace{0.4em}
\subfloat[Processor execution rate of (0.83,0.67,1.0)]{%
\includegraphics[width=0.3\linewidth]{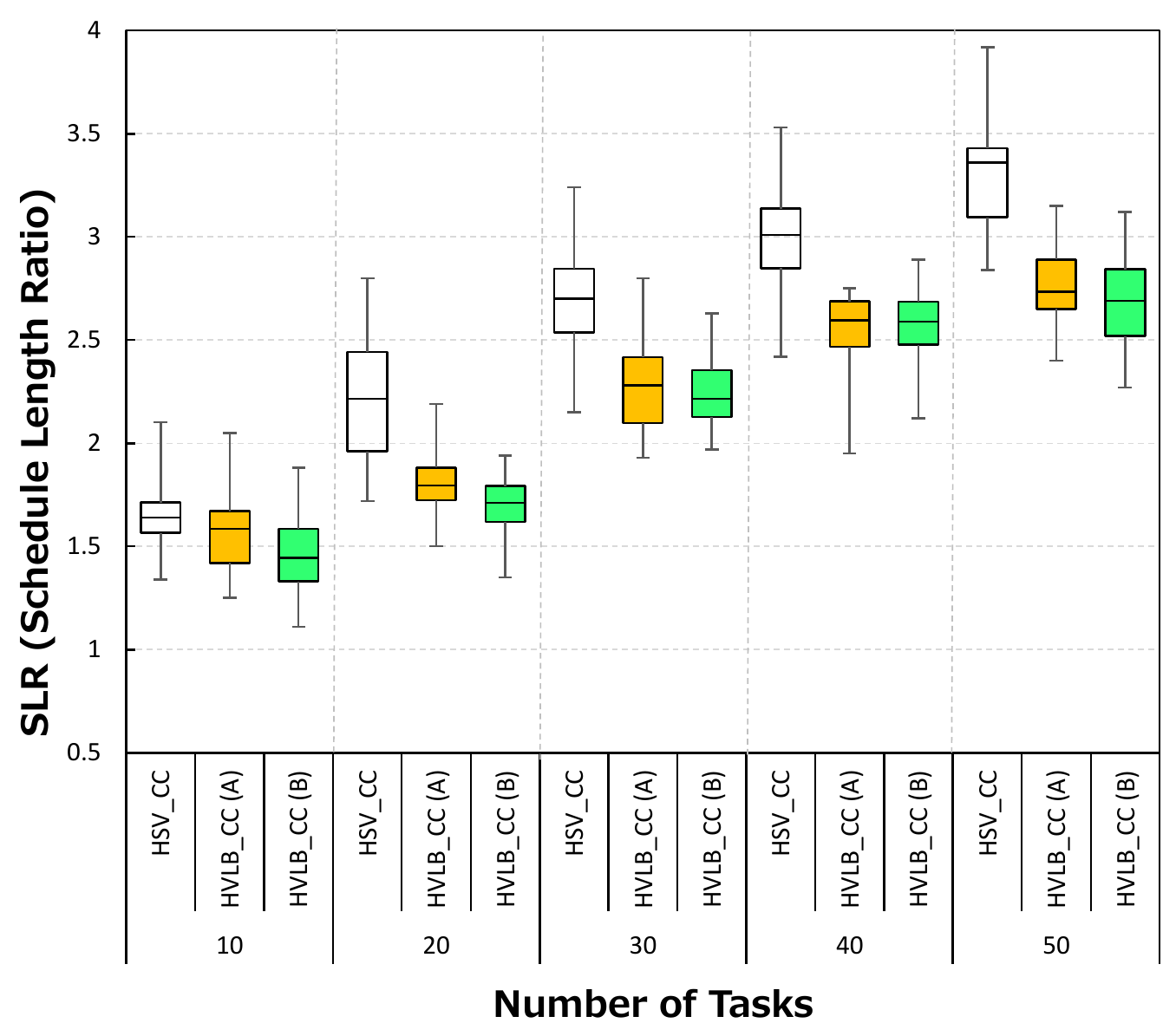}}%
\hspace{0.4em}
\subfloat[Processor execution rate of (0.67,0.83,1.0)]{%
\includegraphics[width=0.3\linewidth]{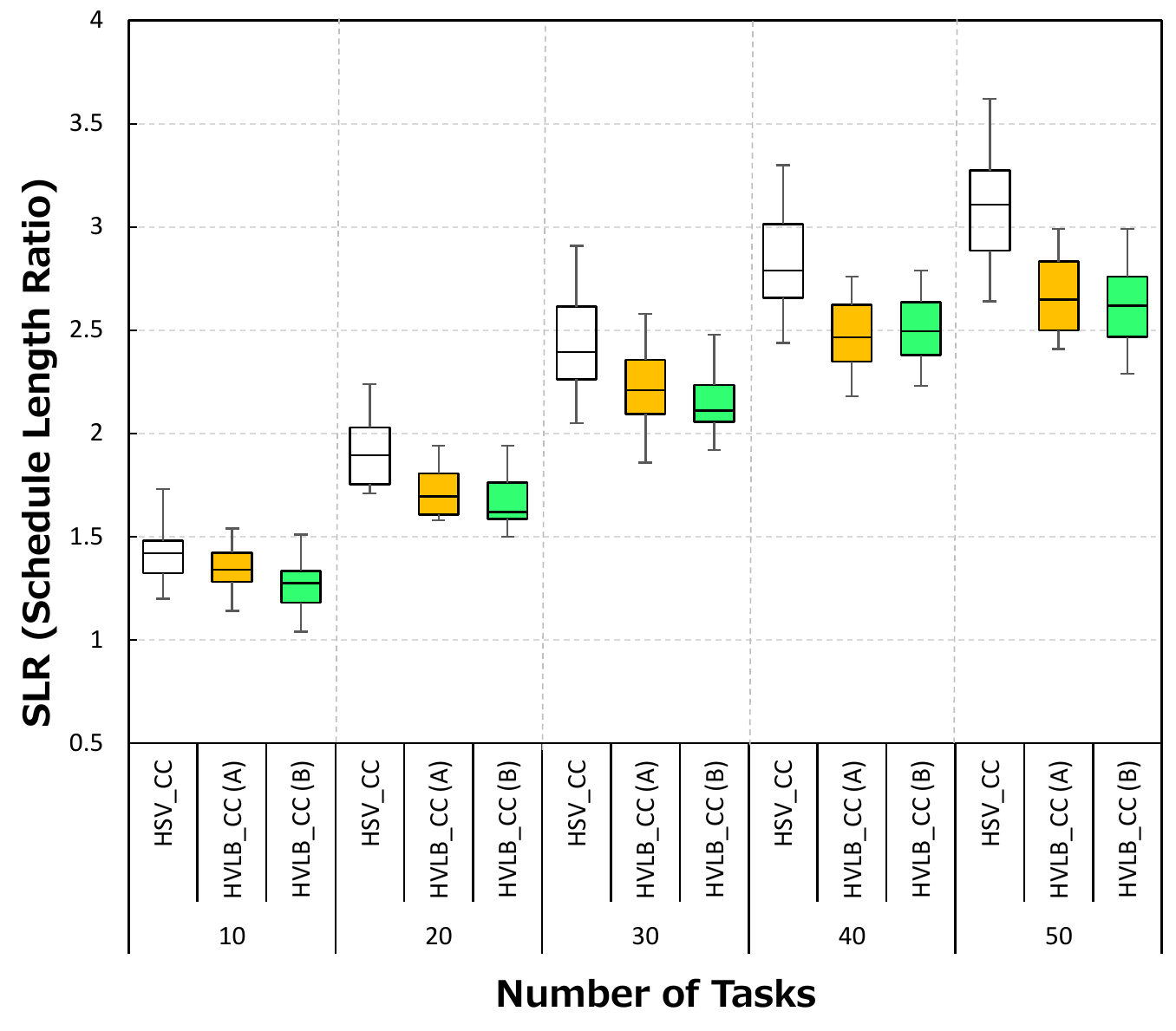}}%
\\
\subfloat[processor execution rate of (1.0,0.83,0.67)]{%
\includegraphics[width=0.3\linewidth]{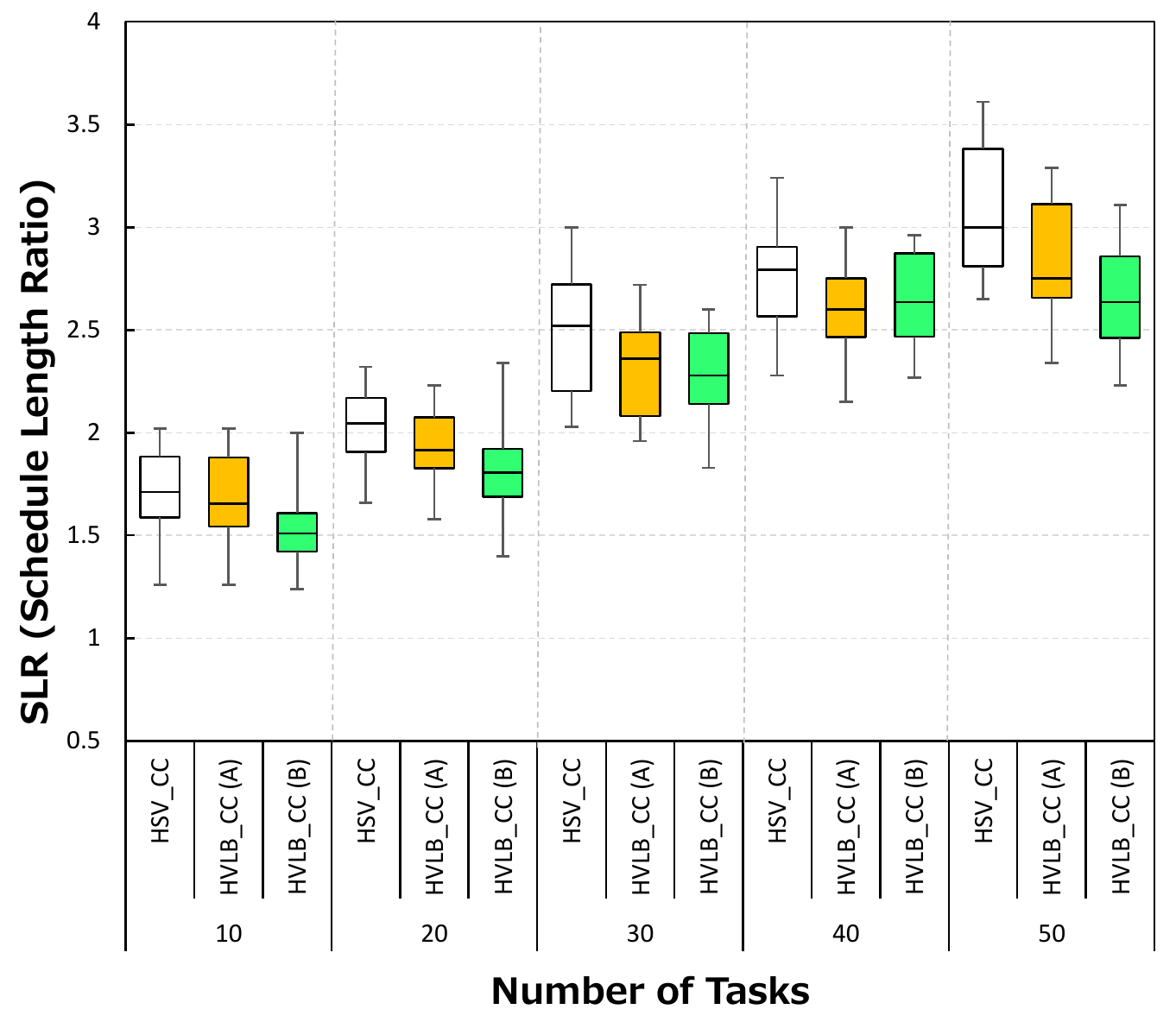}}%
\hspace{0.4em}
\subfloat[Processor execution rate of (0.83,1.0,0.67)]{%
\includegraphics[width=0.3\linewidth]{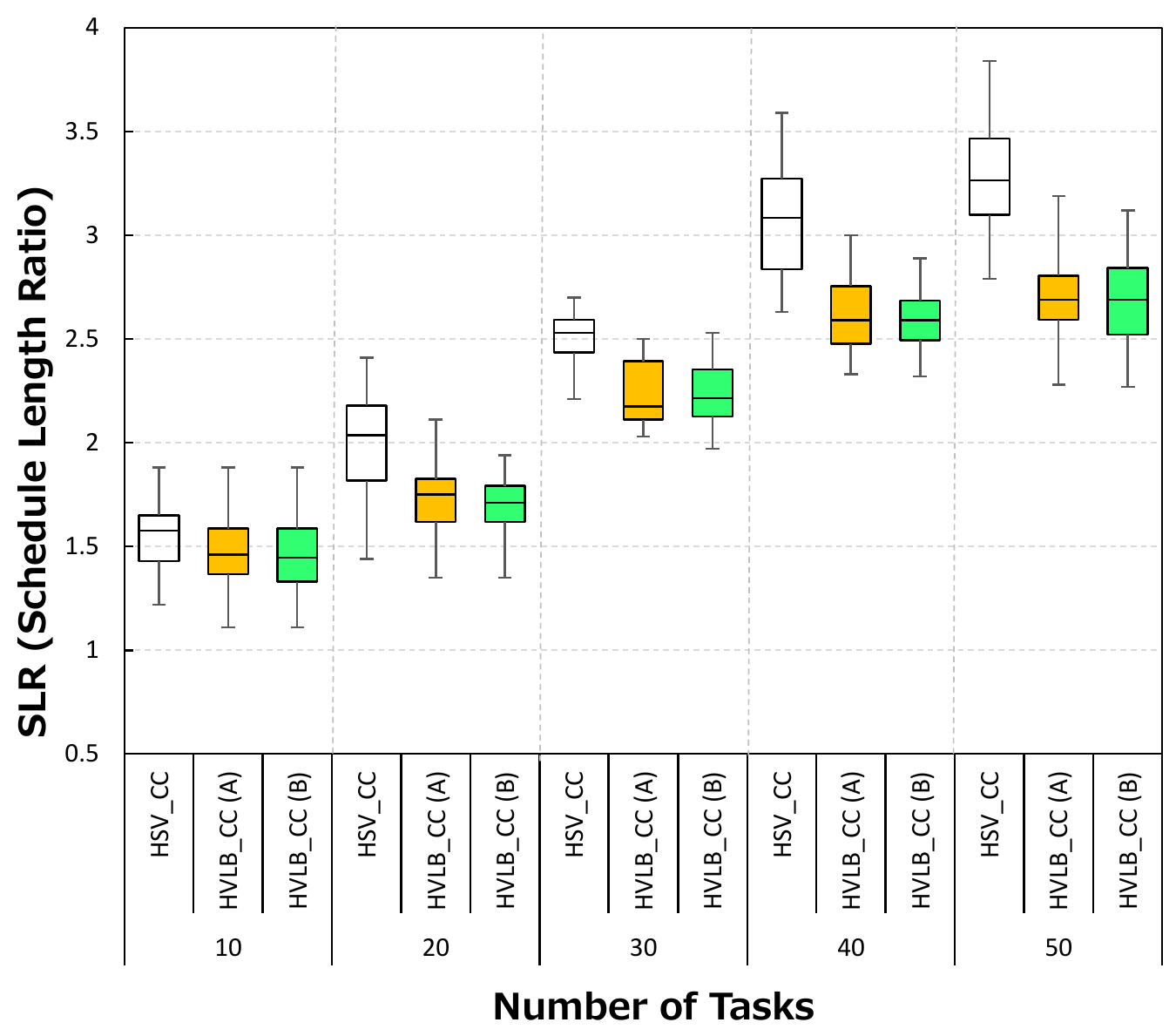}}%
\hspace{0.4em}
\subfloat[Processor execution rate of (0.67,1.0,0.83)]{%
\includegraphics[width=0.3\linewidth]{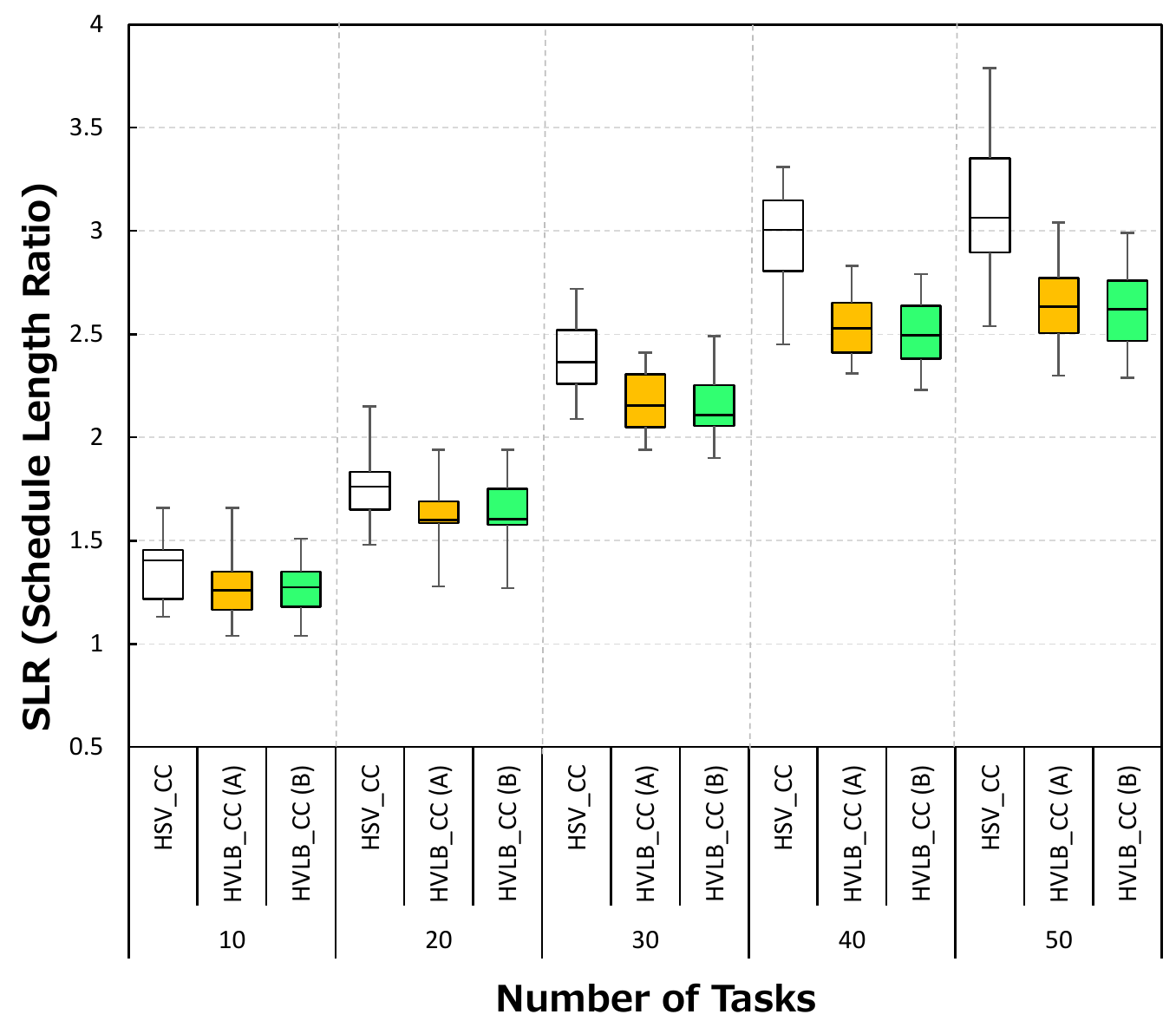}}%
\\
\subfloat[Processor execution rate of (1.0,0.67,0.83)]{%
\includegraphics[width=0.3\linewidth]{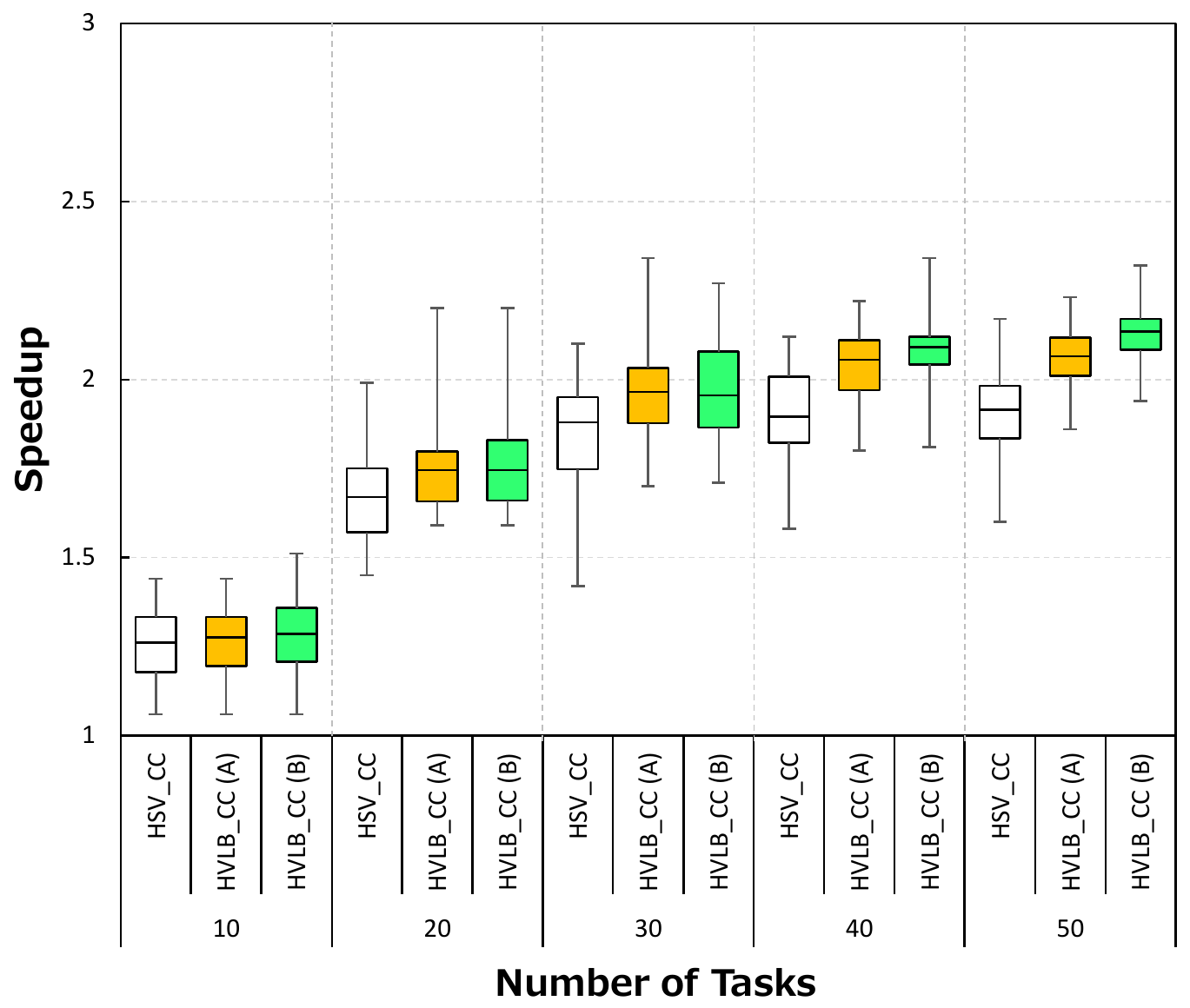}}%
\hspace{0.4em}
\subfloat[Processor execution rate of (0.83,0.67,1.0)]{%
\includegraphics[width=0.3\linewidth]{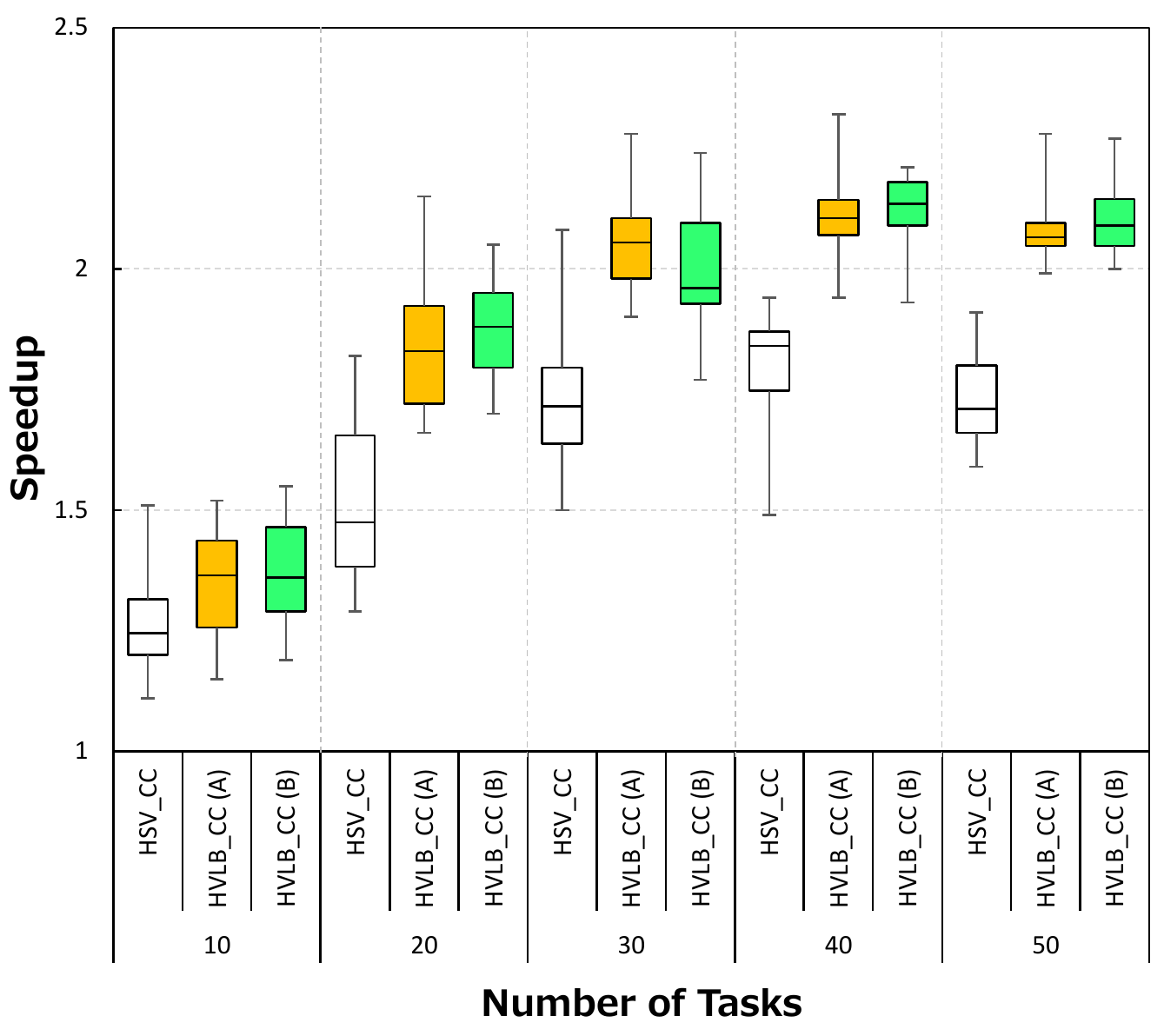}}%
\hspace{0.4em}
\subfloat[Processor execution rate of (0.67,0.83,1.0)]{%
\includegraphics[width=0.3\linewidth]{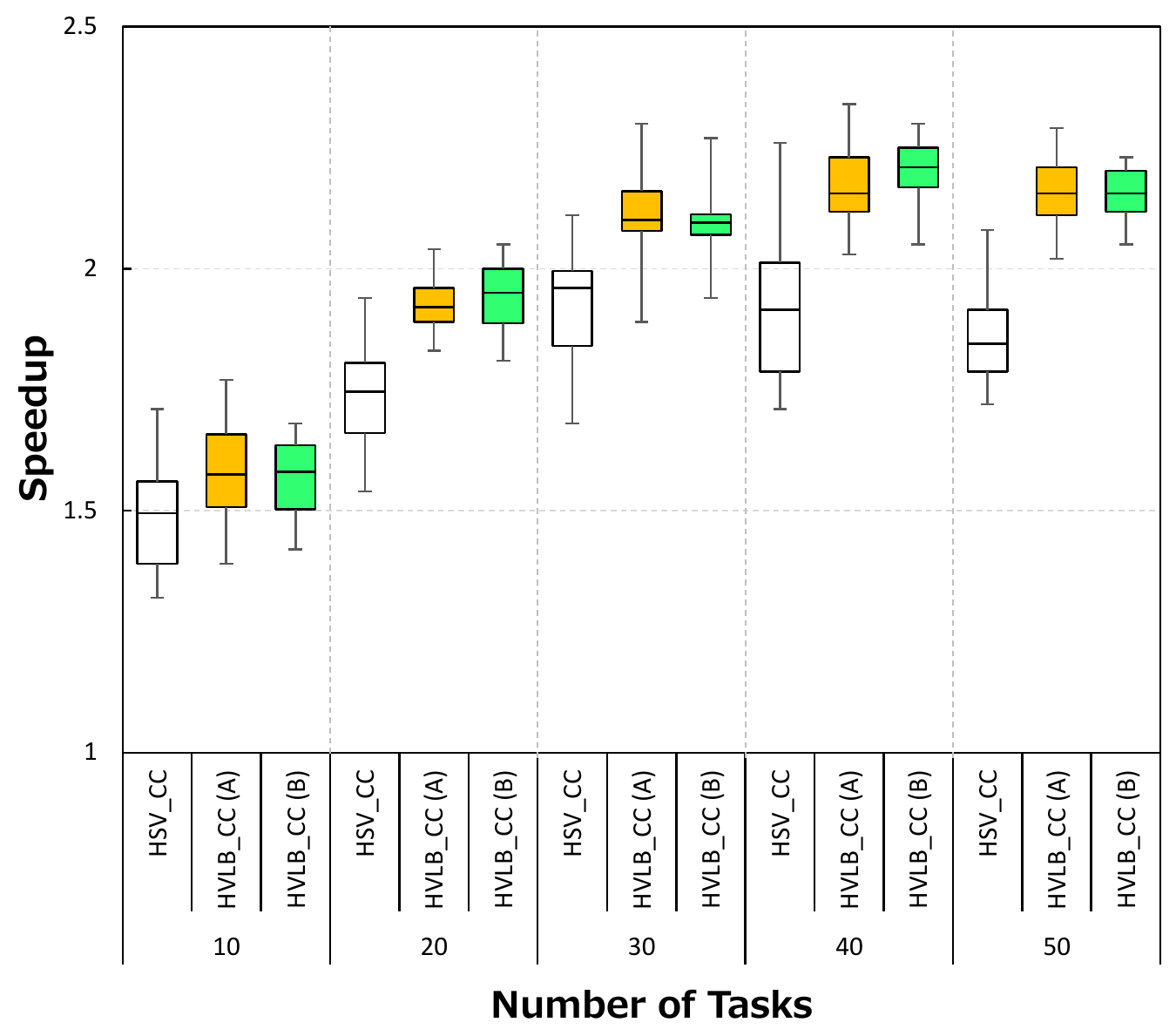}}%
\\

\subfloat[Processor execution rate of (1.0,0.83,0.67)]{%
\includegraphics[width=0.3\linewidth]{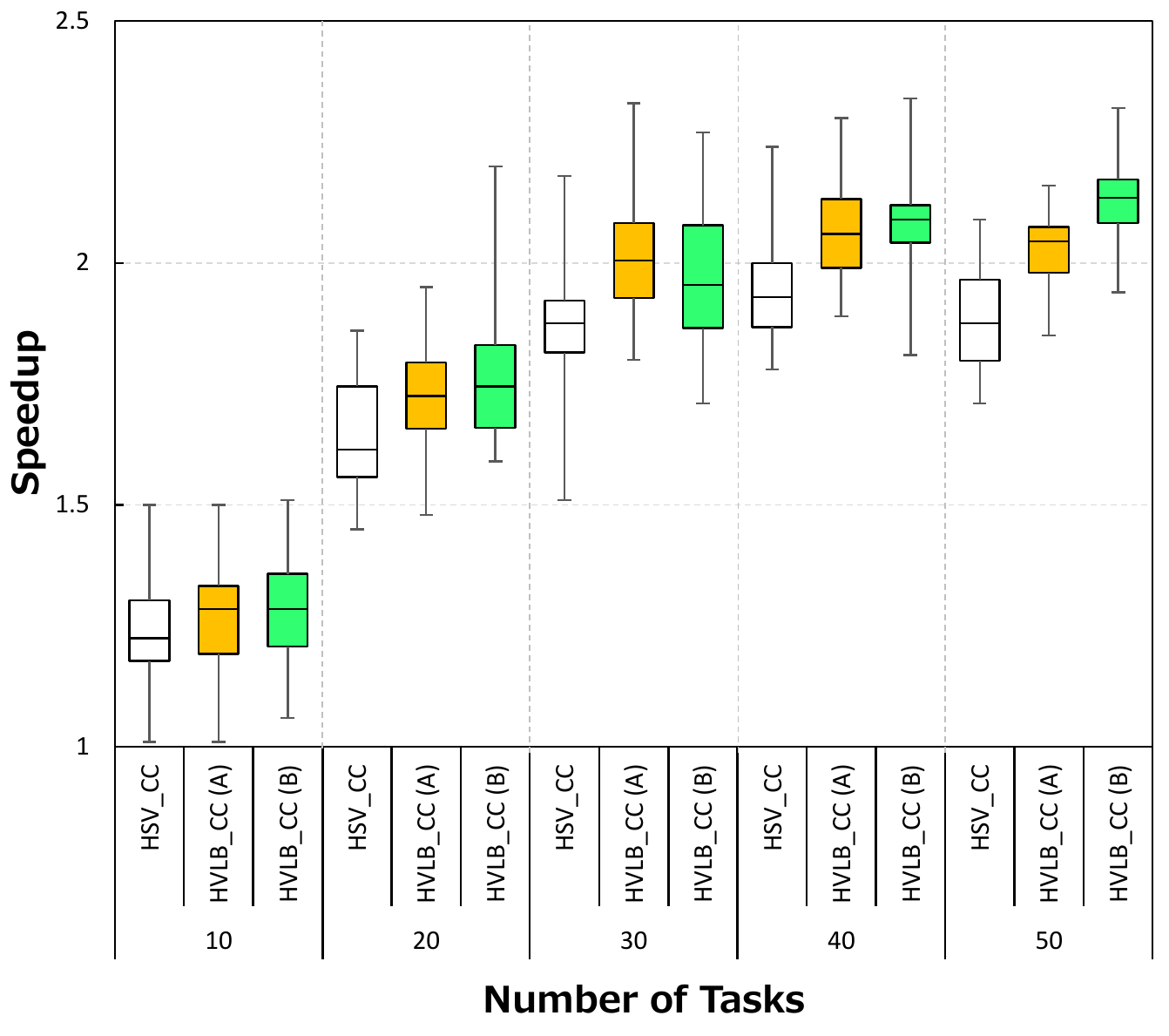}}%
\hspace{0.4em}
\subfloat[Processor execution rate of (0.83,1.0,0.67)]{%
\includegraphics[width=0.3\linewidth]{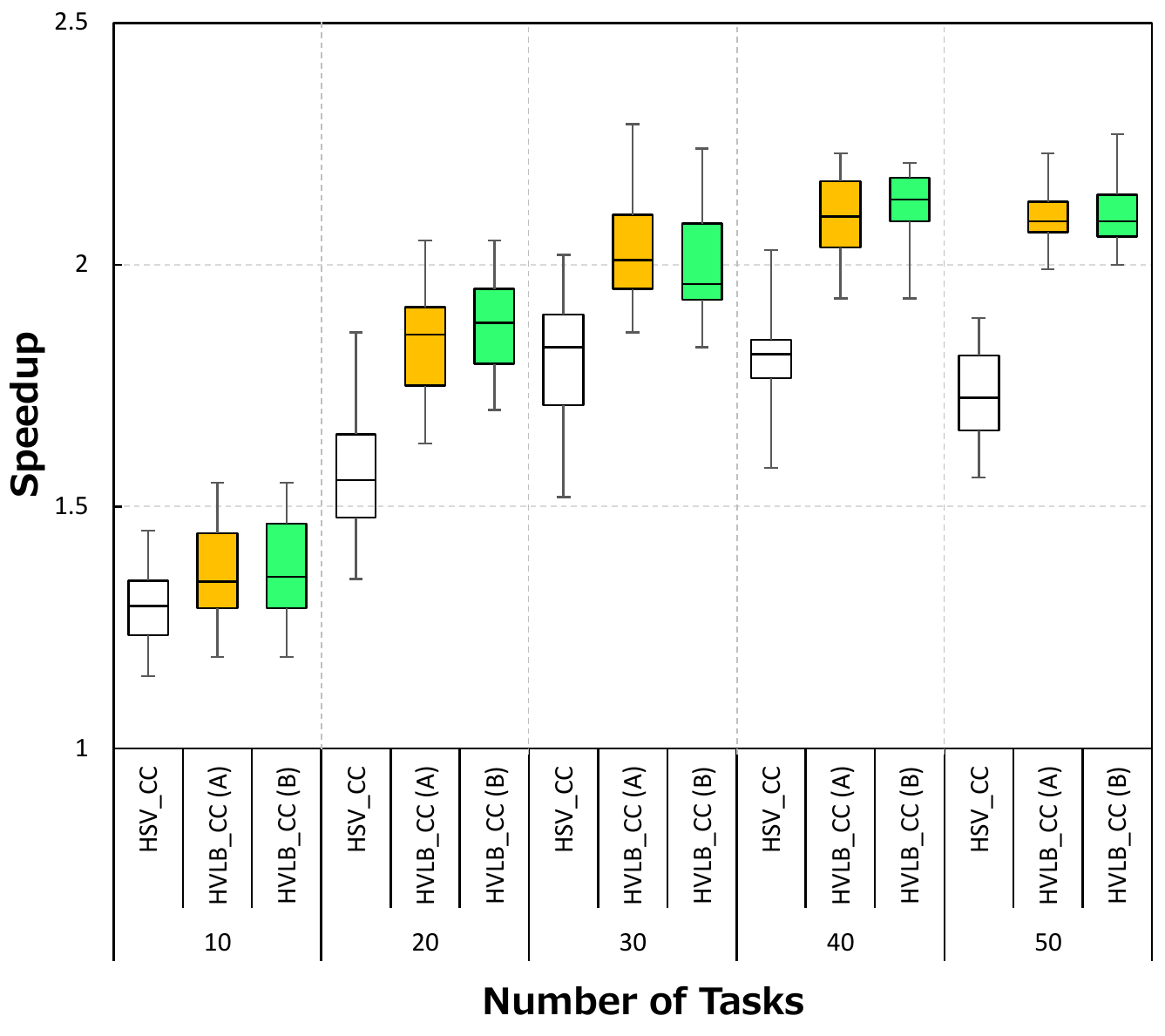}}%
\hspace{0.4em}
\subfloat[Processor execution rate of (0.67,1.0,0.83)]{%
\includegraphics[width=0.3\linewidth]{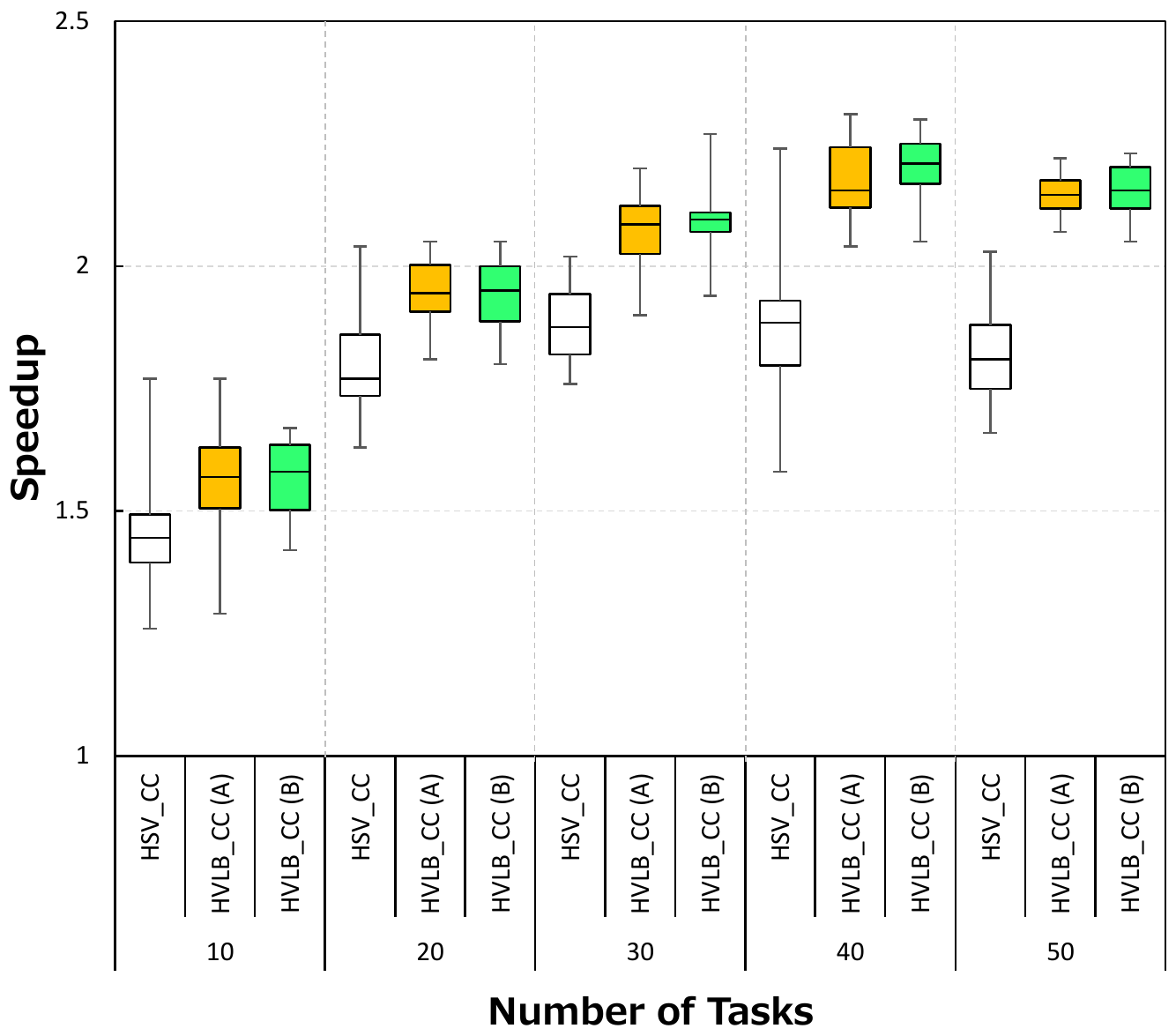}}%

\caption{\color{black} The {\it SLR} and {\it speedup} values of HSV\_CC and HVLB\_CC at the varying processor execution rate and CCR=1.0.}
\label{fig:slrspeed_set}
\end{center}
\vspace{-0.5cm}
\end{figure*}

\begin{figure*}[t]
\begin{center}

\subfloat[Processor execution rate of (1.0,0.67,0.83)]{%
\includegraphics[width=0.3\linewidth]{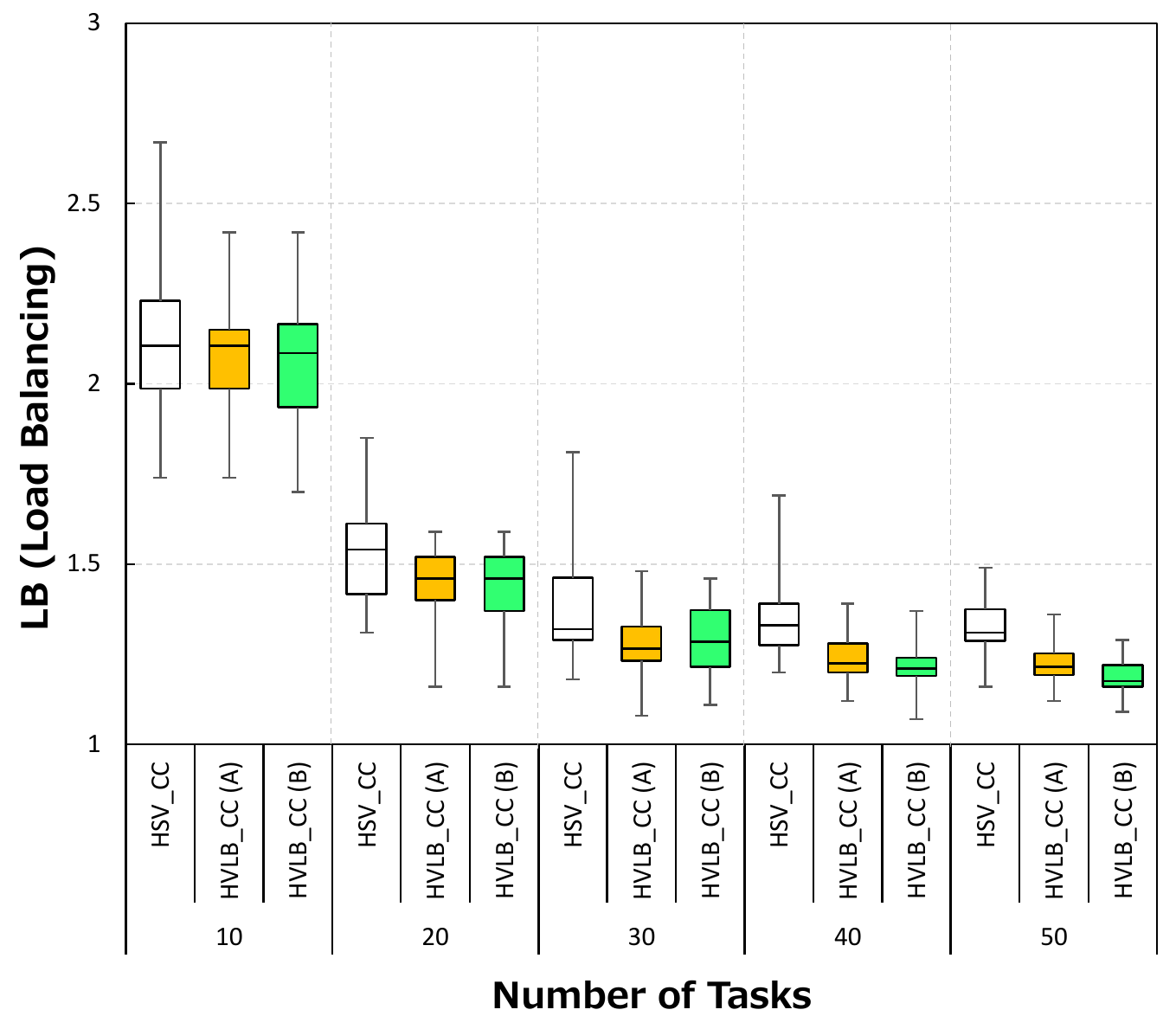}}%
\hspace{1em}
\subfloat[Processor execution rate of (0.83,0.67,1.0)]{%
\includegraphics[width=0.3\linewidth]{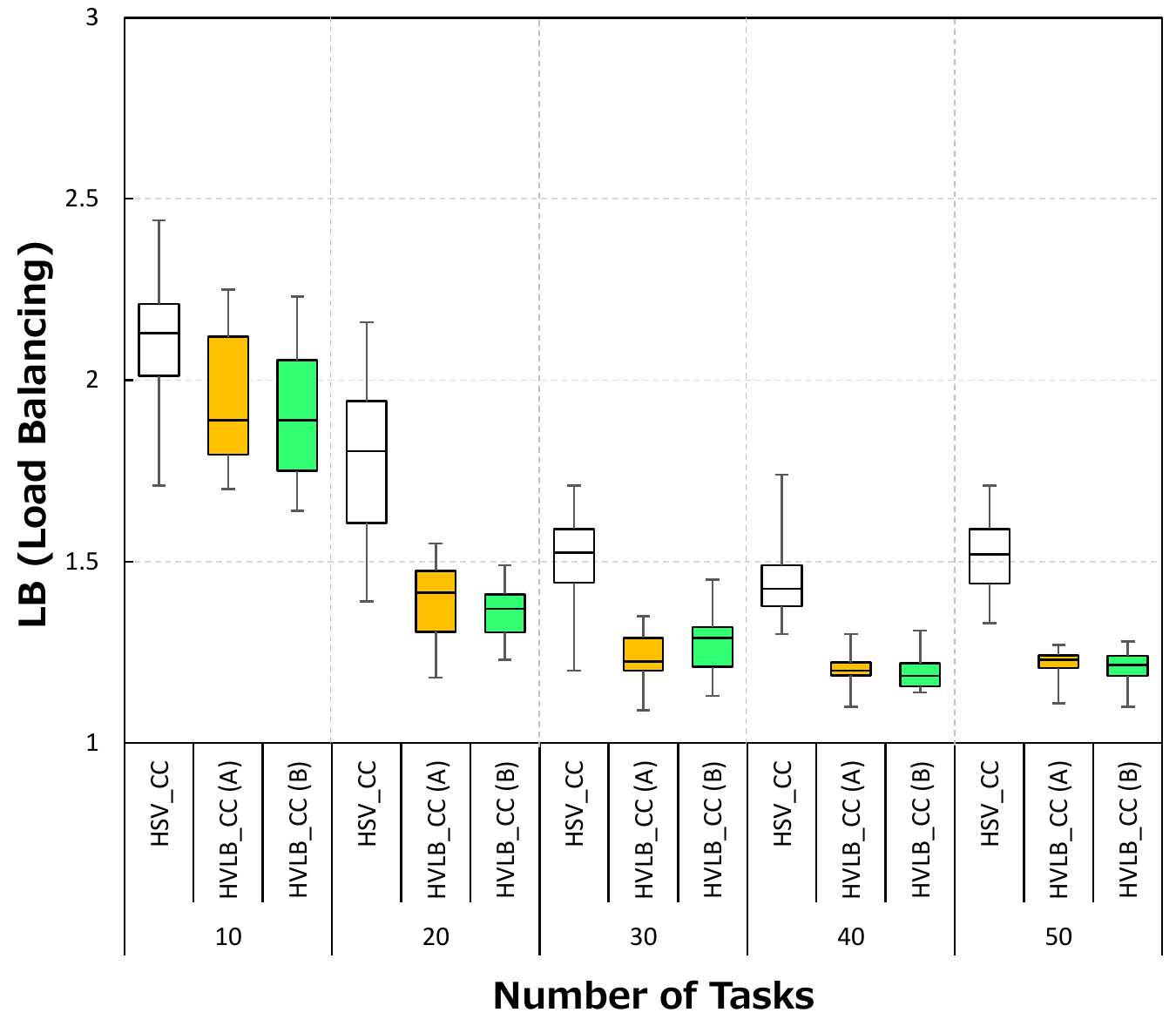}}%
\hspace{1em}
\subfloat[Processor execution rate of (0.67,0.83,1.0)]{%
\includegraphics[width=0.3\linewidth]{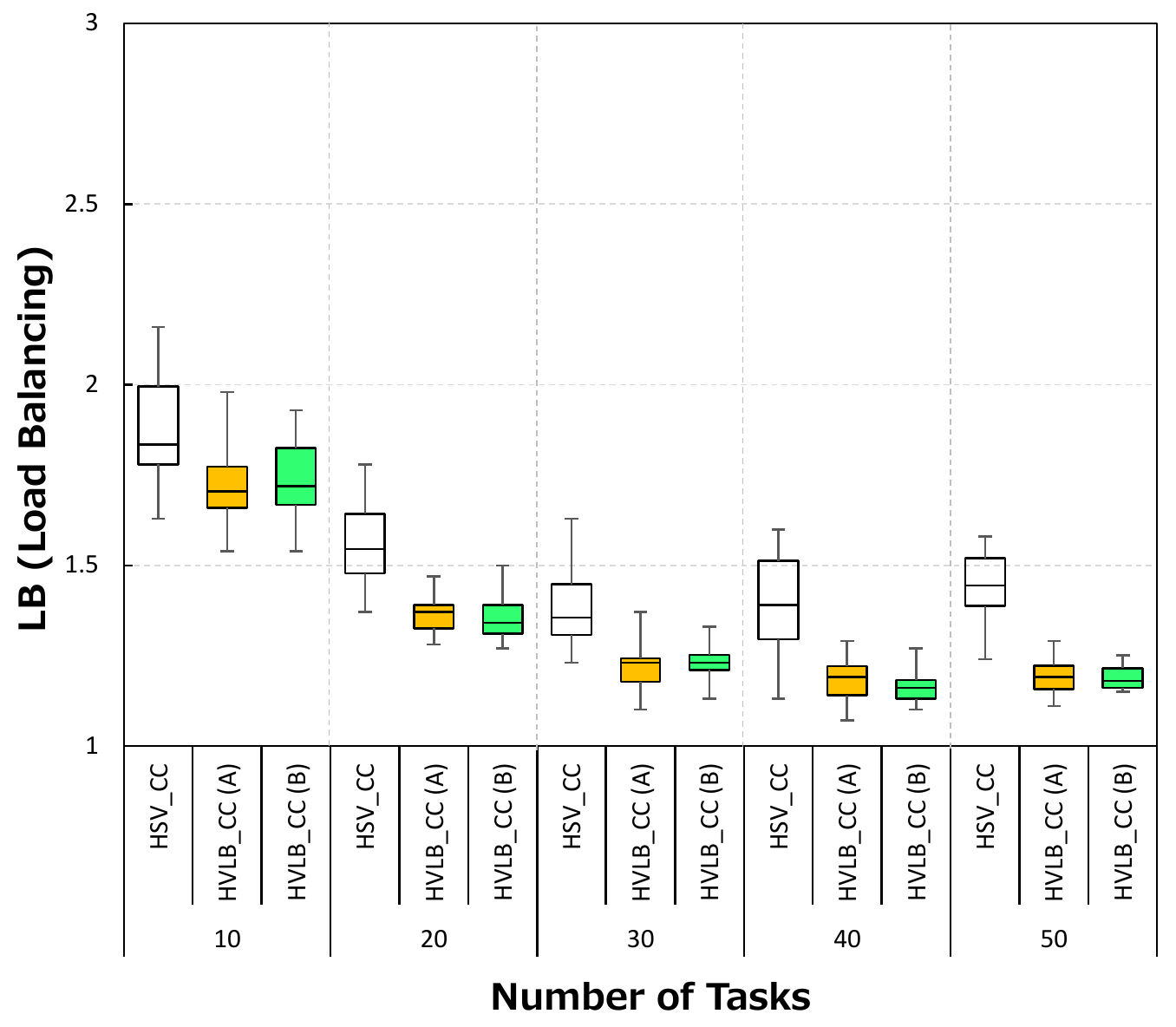}}%
\\
\subfloat[Processor execution rate of (1.0,0.83,0.67)]{%
\includegraphics[width=0.3\linewidth]{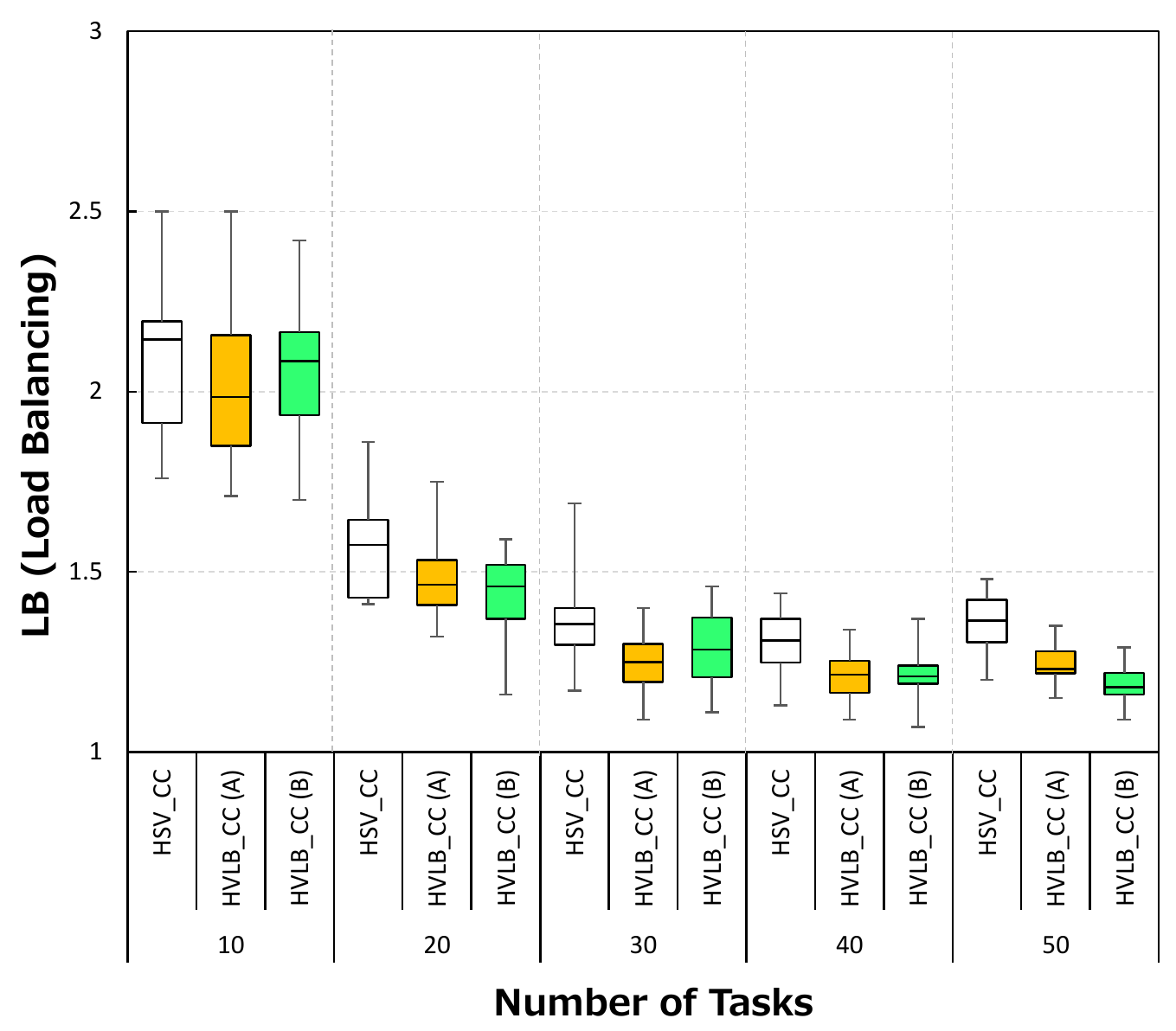}}%
\hspace{1em}
\subfloat[Processor execution rate of (0.83,1.0,0.67)]{%
\includegraphics[width=0.3\linewidth]{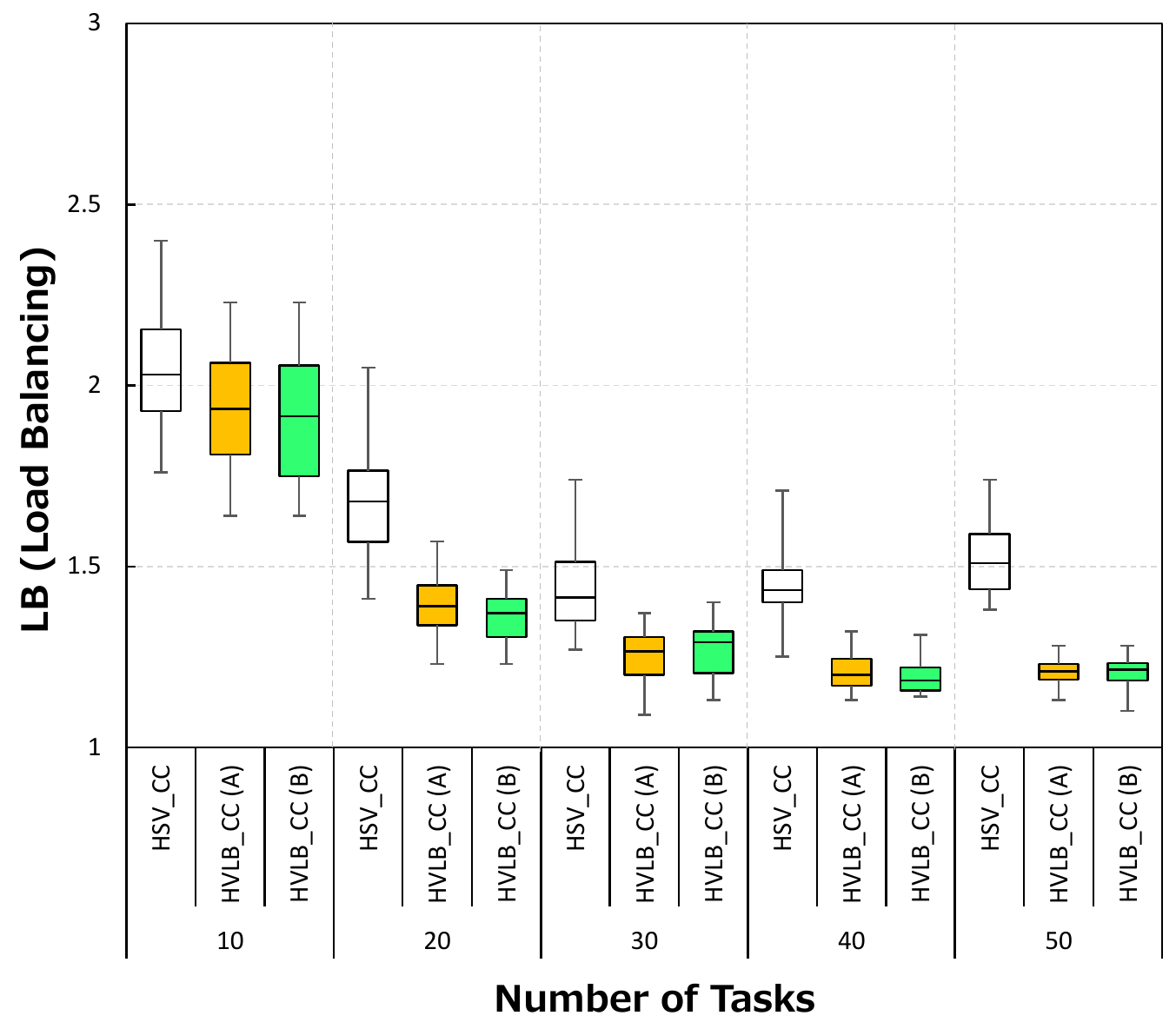}}%
\hspace{1em}
\subfloat[Processor execution rate of (0.67,1.0,0.83)]{%
\includegraphics[width=0.3\linewidth]{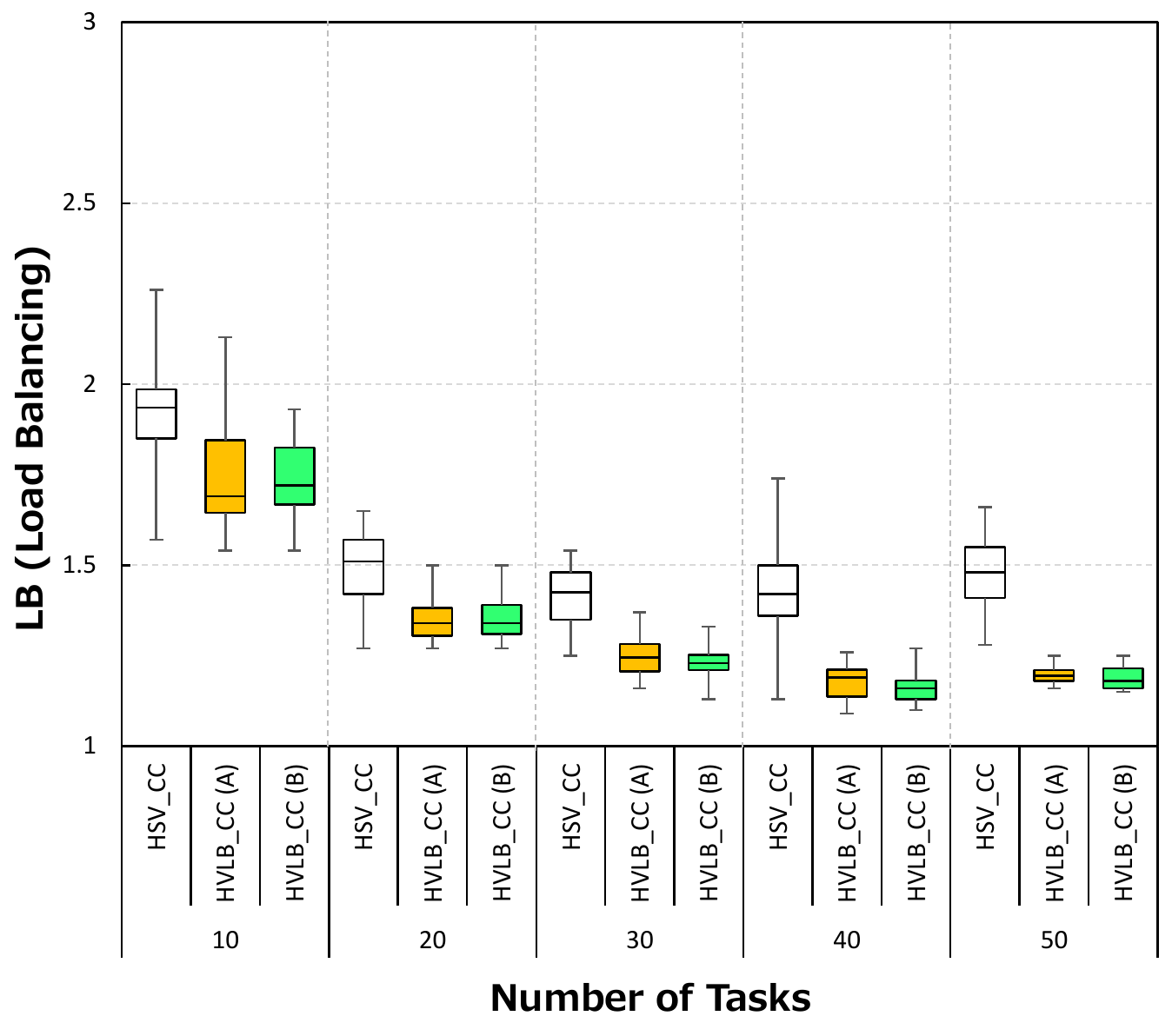}}%
\caption{\color{black}The Load Balancing values of HSV\_CC vs HVLB\_CC with increasing the number of taskss.}
\label{fig:load_set}
\end{center}
\vspace{-0.5cm}
\end{figure*}

\subsection{Experiment 1: Comparison of Scheduling Accuracy and Performance}
Here, we discuss the {\it SLR} and {\it speedup} results for a varying number of tasks to verify the superiority of the proposed algorithms in terms of scheduling accuracy and performance. For each number of tasks, we randomly generated 100 task graphs with variable computation times according to the execution times of the three processors. Note that {\it CCR} was set to 1. The number of tasks increased from 10 to 50 at increments of 10. These numbers were determined based on a real stream processing case in an automotive DSMS. In each task graph, $\alpha$, which can adjust the influence of LB on the processor selecting phase, varies from 0 to 20 in 0.01 increments. The scheduling results that could minimize {\it makespan} by increasing $\alpha$ were chosen in the proposed HVLB\_CC (A) and HVLB\_CC (B) algorithms. \textbf{Figure} \ref{fig:slrspeed_set} shows the comparative results for {\it SLR} and {\it speedup}.

We first present the experimental results for {\it SLR} that were obtained by HSV\_CC and the proposed HVLB\_CC (A) and HVLB\_CC (B) algorithms with three processors. As shown in \textbf{Figures} \ref{fig:slrspeed_set} { \color{black}(a)-(f)}, HVLB\_CC (A) and HVLB\_CC (B) always outperform HSV\_CC in terms of  {\it SLR} in the three processor execution time rate patterns. The worst values of {\it SLR} obtained using HSV\_CC are { \color{black}3.49, 3.92, 3.62, 3.61, 3.84 and 3.79}, corresponding to processor execution rates of { \color{black}$[1.0, 0.67, 0.83]$, $[0.83, 0.67, 1.0]$,  $[0.67, 0.83, 1.0]$, $[1.0,0.83,0.67]$,$[0.83,1.0,0.67]$, and $[0.67,1.0,0.83]$ }respectively. When the processor execution rate is { \color{black}$[1.0, 0.67, 0.83]$}, {\it SLR} is the lowest value because the tasks and messages are scheduled on high-capability processors and communication links. Thus, {\it makespan} can be decreased because the tasks are executed in less time due to the high-capability processors. When the processor execution rates are { \color{black}$[0.83, 0.67, 1.0]$}, the tasks are likely to be assigned to $p_1$ because it is the higher-capability processor. In these two processor patterns, the tasks and messages are nearly equally assigned to the three processors; however, {\it makespan} increases because some tasks are executed to the lowest-capability processor.

The worst {\it SLR} values obtained using HVLB\_CC (A) and HVLB\_CC (B) are { \color{black}3.06, 3,15, 2.99, 3.29, 3.04 and 3.19}, and { \color{black}{3.11, 3.12, 2.99, 3.11, 2.99 and 3.12}}, corresponding to processor execution rates of { \color{black}$[1.0, 0.67, 0.83]$, $[0.83, 0.67, 1.0]$,  $[0.67, 0.83, 1.0]$, $[1.0,0.83,0.67]$,$[0.83,1.0,0.67]$, and $[0.67,1.0,0.83]$ }, respectively. These results show that the proposed scheduling algorithms outperform HSV\_CC in term of worst-case SLR. In HSV\_CC, the processor selection criteria are based on {\it EFT} and {\it LDET} which are largely influenced by the processor execution rate. Thus, the tasks and messages are intended to be allocated to specific processors and communication links. However, in the proposed HVLB\_CC (A) and (B) algorithms, allocating tasks and messages to specific processors and links can be avoided because, with $\alpha$, considering and balancing the processor loads during the processor selection phase can affect the schedule length positively. 
{ \color{black}
\textbf{Figures} \ref{fig:slrspeed_set} { \color{black}(h)-(m)} show the results for {\it speedup} with varying number of tasks that were obtained using HSV\_CC and the proposed HVLB\_CC (A) and (B) algorithms. The best {\it speedup} value is 2.26 when the processor execution rates are $[0.67,0.83,1.0]$ for HSV\_CC. For HVLB\_CC (A) and (B), the best {\it speedup} values are 2.34 and 2.34 when the processor execution rates are $[1.0,0.67,0.83]$, and the proposed algorithms outperform HSV\_CC for all numbers of tasks and all processor execution rates because, as mentioned above, HVLB\_CC (A) and (B) schedule tasks by considering the processor loads and minimizing {\it makespan} by adjusting $\alpha$. 
}

\subsection{Experiment 2: Comparison of Load Balance over Processors}

Here, we analyze the LB results with a varying number of tasks to confirm the load balance between the processors. For each number of tasks, we randomly generated 100 task graphs with variable computation times according to the execution times of the three processors. In addition, {\it CCR} was set to 1.0. The number of tasks was increased from 10 to 50 in increments of 10, and, in each task graph, $\alpha$ was selected to minimize {\it makespan}.

As shown in \textbf{Figure} \ref{fig:load_set}, the proposed HVLB\_CC (A) and (B) algorithms show better performance in terms of LB with a varying number of tasks and for all of the processor execution rates, because the proposed algorithms consider both of the processor load in the processor selection phase and the factors considered by HSV\_CC (i.e., {\it EFT} and {\it LDET}). In HSV\_CC, tasks and messages are intended to be allocated to specific processors and communication links because both {\it EFT} and {\it LDET} are highly dependent on the processor execution rate. However, in the proposed HVLB\_CC (A) and (B) algorithms, allocating tasks and messages to higher-speed processors and links can be avoided. The proposed algorithms yield balanced processor loads, and can decrease {\it makespan} by { \color{black}utilizing processors and links that are not sufficiently used in HSV\_CC}. 

\subsection{Experiment 3: Comparison of Scheduling Accuracy with Varing Communication-to-Computation Ratio}
In this experiment, we observed {\it SLR} while varying {\it CCR}. The number of tasks was fixed to 20. The processor execution rates of the three processors were set to 0.83, 1.0, 0.67. We randomly generated 100 task graphs with variable computation times according to the execution time rate of three processors, and {\it CCR} was varied from 0.1 to 10. 

\begin{figure}
\centering
\includegraphics[width=0.9\linewidth]{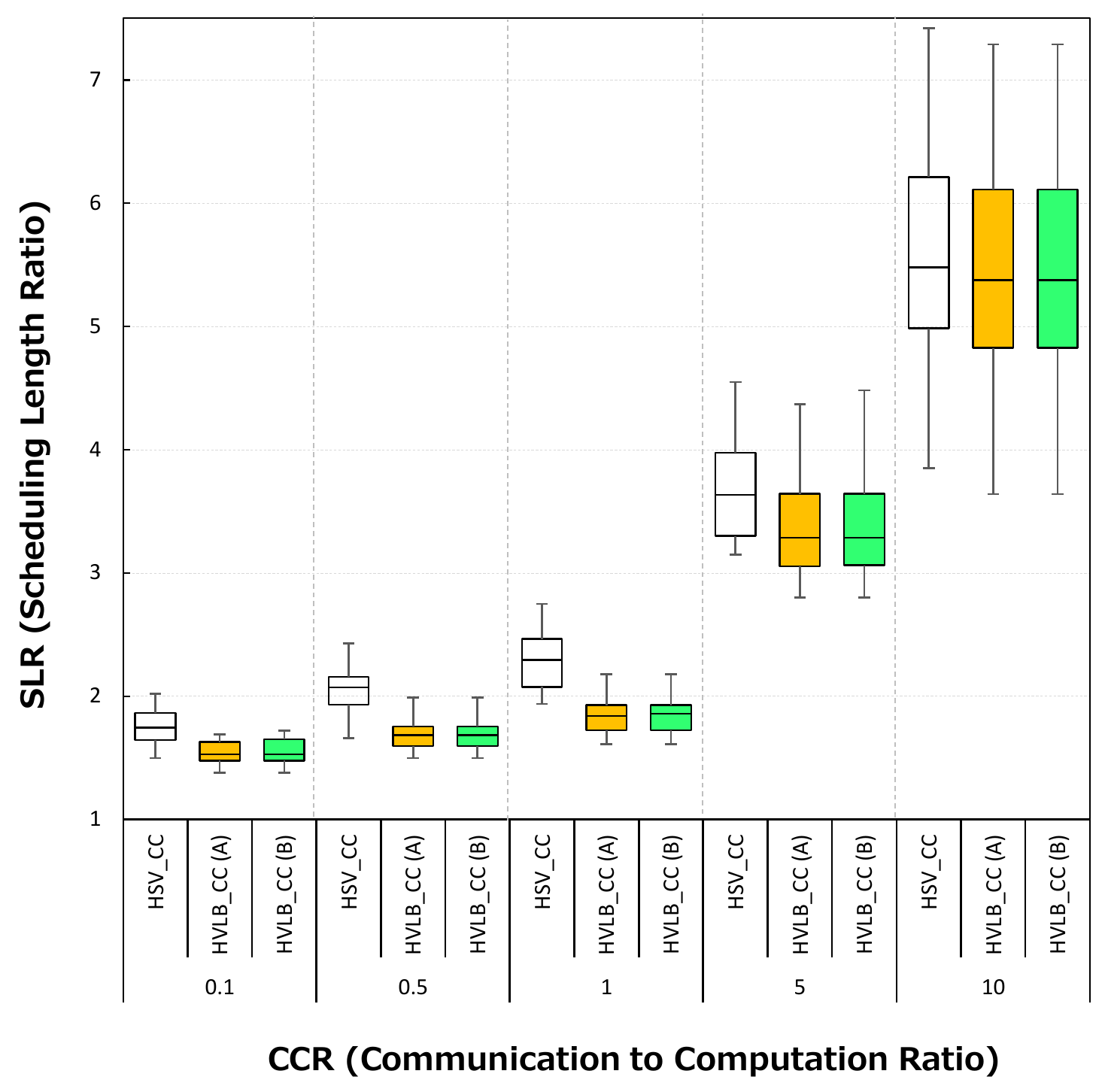}
\caption{{\it SLR} values for HSV\_CC and HVLB\_CC with increasing CCR}
\label{fig:ccr_result_2}
\vspace{-0.3cm}
\end{figure}

\textbf{Figure} \ref{fig:ccr_result_2} shows the {\it SLR} values for HSV\_CC, HVLB\_CC (A), and HVLB\_CC (B) with different {\it CCR} values. 

For {\it CCR} = 0.1 and 0.5, the communication time is of low significance compared to the computation time (i.e., these problems are computation-intensive). The worst-case {\it SLR} values of the proposed HVLB\_CC algorithms are 16.3 \% and 18.1 \% better than that of HVC\_CC when {\it CCR} is 0.1 and 0.5, respectively. Similarly, the best-case {\it SLR} values of the proposed HVLB\_CC algorithms are 8 \% and 9.6 \% better than that of HVC\_CC when {\it CCR} is 0.1 and 0.5, respectively. For {\it CCR} = 1, the {\it SLR} values of HVLB\_CC (A) and HVLB\_CC (B) are the same. Compared to the {\it SLR} values of HSV\_CC, the proposed algorithms can improve by 20.7 \% and 17 \% in the worst and best-case scenarios, respectively. For {\it CCR} = 5 and 10, the tasks in the task graphs have longer data transfer times than computation times. For the worst-case {\it SLR}, HVLB\_CC (A) can improve by 3.9 \% and 1.75 \% compared to HSV\_CC when {\it CCR} is 5 and 10, respectively. Similarly, HVLB\_CC (B) can improve by 1.54 \% and 1.75 \% compared to HSV\_CC when {\it CCR} is 5 and 10, respectively. For the best-case {\it SLR}, HVLB\_CC (A) and (B) can improve by 11.1 \% and 5.5 \% when {\it CCR} is 5 and 10, respectively. Note that the gap between HSV\_CC and HVLB\_CC relative to {\it SLR} decreases when {\it CCR} becomes greater than 1 because tasks tend to be assigned to high-capability processors first when {\it CCR} is greater than 1, and successor tasks are not mostly assigned to different processor with a precedence task. A higher {\it CCR} value increases the communication time more than the computation time, which causes {\it EFT} to increase for the task. 

It can be observed that the improvement becomes significantly greater when {\it CCR} is decreased from 10 to 1 and when {\it CCR} is increased from 0.1 to 1. The greatest improvements with the proposed algorithm are possible with a {\it CCR} value of 1. Note that very little improvement is observed when {\it CCR} is 10. As a result, the HVLB\_CC (A) and (B) algorithms can consider the heterogeneity of both computation and communication for task ordering.
 
%The count decreases as CCR increases because the idle time gaps are larger for higher CCR

\subsection{Experiment 4: Comparison of Scheduling Failure Rate}
As mentioned in Section \ref{sec:problem2}, HSV\_CC is restricted in scheduling for various DAGs. In this experiment, we confirmed that how much the rate could be improved by considering the depth factor in a DAG with the proposed HVLB\_CC (B) algorithm. 

We generated 1,000 random task graphs. In the stream processing system, an operator, such as {\it Join} or {\it Union}, has two in- and out-degrees, and a user-defined operator can have a maximum in-degree of three. Thus, we generated random DAGs whose maximum in-degrees were 2 and out-degrees were 3. To confirm successful scheduling for HSV\_CC and HVLB\_CC (B), the scheduling failure rate ({\it SFR}) is defined as follows:

\begin{equation}
\label{eq:schedule_failure}
\hspace*{-7mm}
  {\it SFR} = \\\frac{{\it number\ of\ failed\ DAGs}}{{\it total\ number\ of\ requested\ DAGs}} \times 100
\end{equation}

\begin{figure}
\centering
\includegraphics[width=1\linewidth]{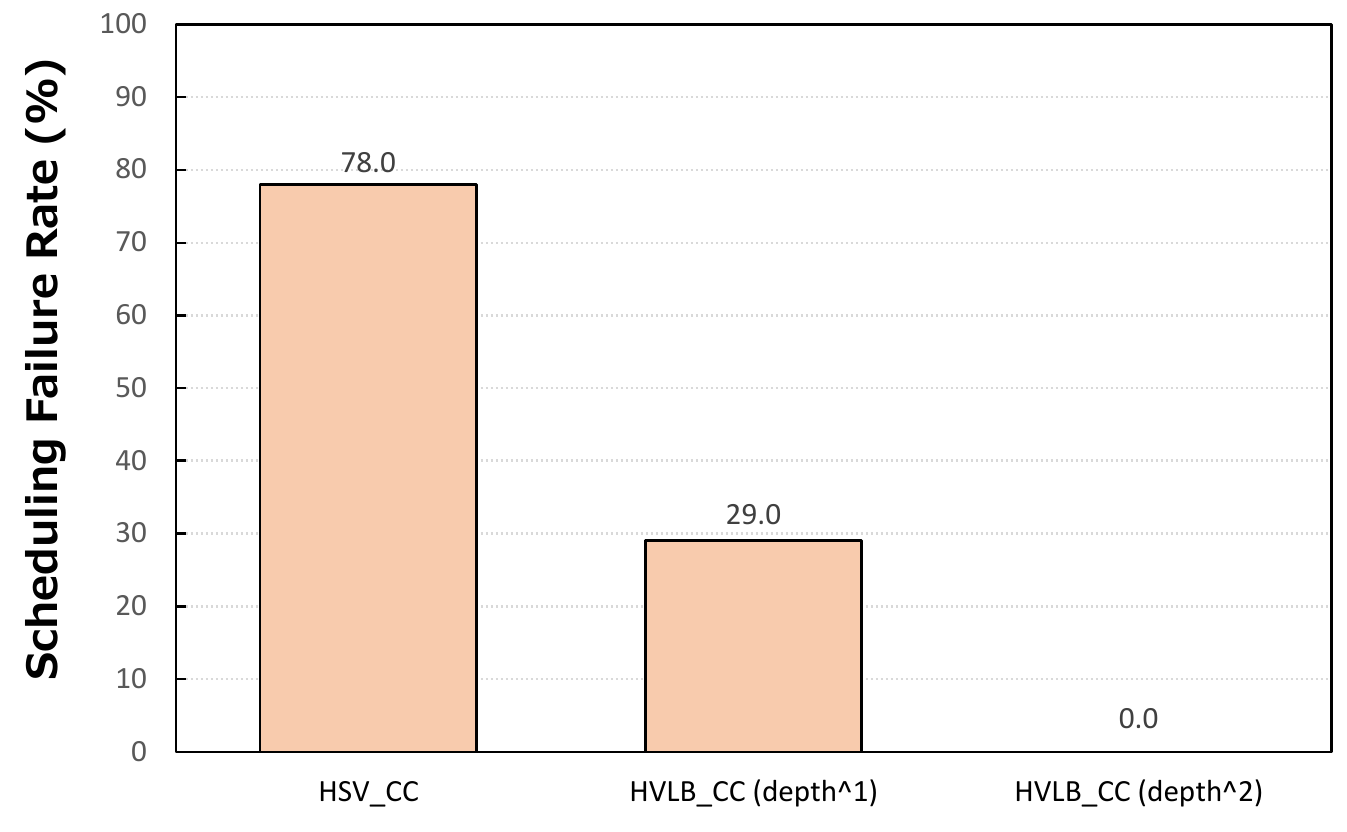}
\caption{{\it SFR}s of HSV\_CC and HVLB\_CC}
\label{fig:schedule_failure}
\vspace{-0.3cm}
\end{figure}

\textbf{Figure} \ref{fig:schedule_failure} shows the {\it SFR}s of HSV\_CC, HVLB\_CC ($depth$), and HVLB\_CC ($depth^2$). The {\it SFR} of HSV\_CC is 78 \%, and the {\it SFT} values of HVLB\_CC ($depth$) and HVLB\_CC ($depth^2$) are 29 \% and 0 \%, respectively. As a result, HSV\_CC is very restricted when the DAGs are random, and it can only schedule perfectly for DAGs with $outd(n_p)\geq outd(n_s)$, where $n_p$ and $n_s$ are predecessor and successor tasks, respectively. We observed the change in {\it SFR} while increasing the impact of the depth factor. When we use Eq. \ref{eq:hprv2} with $depth$, there are still DAGs that experience scheduling failure. In contrast, when we increase the depth factor as $depth^2$ in Eq. \ref{eq:hprv2}, all DAGs are scheduled successfully. Thus, we select $depth^2$ in Eq. \ref{eq:hprv2}.

\subsection{Experiment 5: Comparison of Data Precision with Increasing Input Arrival Rate}

\begin{figure}
\centering
\includegraphics[width=1.0\linewidth]{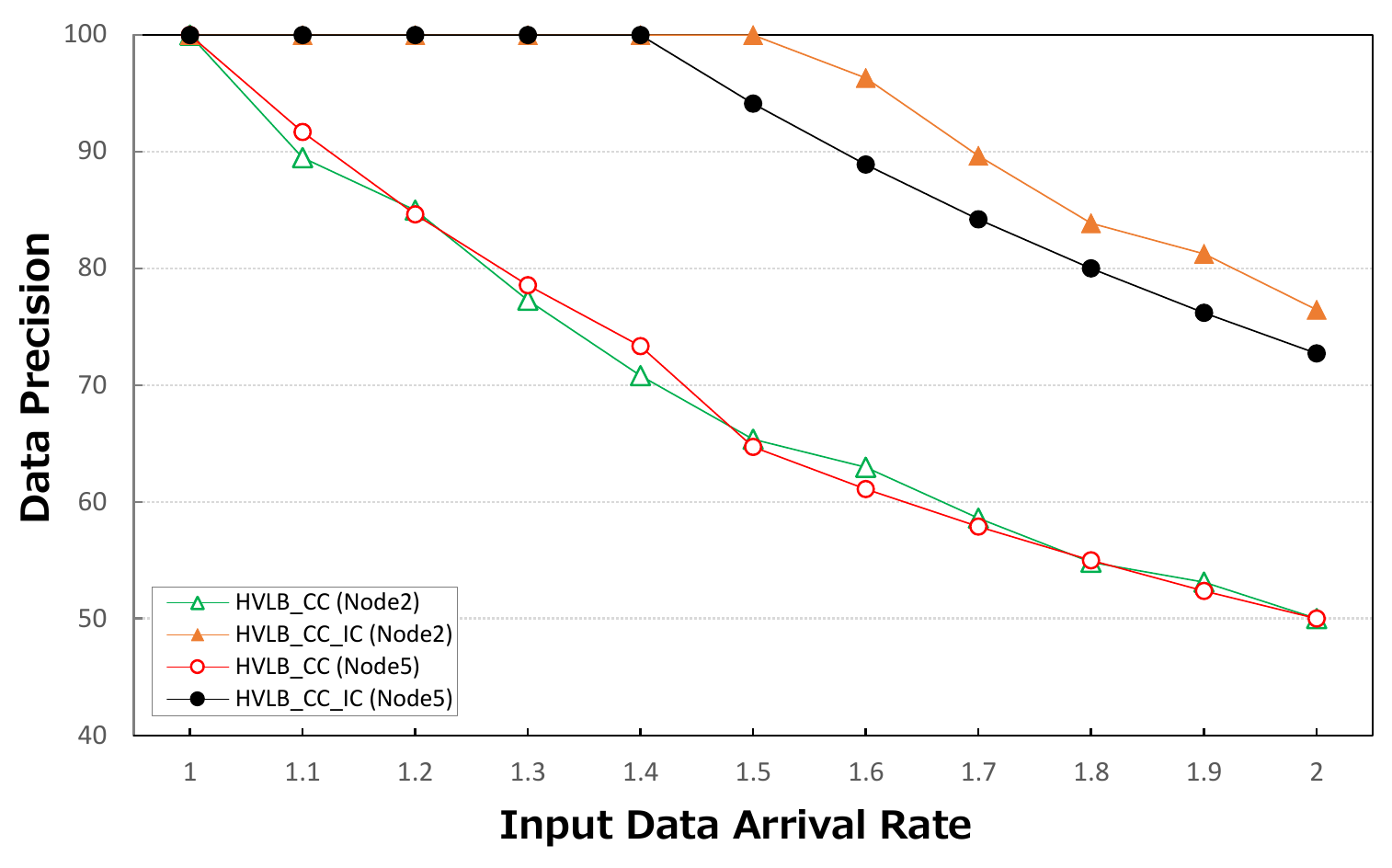}
\caption{Effect of imprecise computation model}
\label{fig:imprecise}
\vspace{-0.3cm}
\end{figure}

\begin{table}[!t]
\begin{center}
\centering
\caption{Computation time matrix used in {\it Experiment 5}}
\label{tab:comp_time2}
\scriptsize
\begin{tabular}{p{1.5cm}p{1.3cm}p{1.3cm}p{0.7cm}} \hline
Task&$p_1$&$p_2$&$p_3$ \\[0.7ex] \hline
$n_1$&26&17&20\\
$n_2$&26&17&20\\
$n_3$&14&9&11\\
$n_4$&12&8&10\\
$n_5$&17&11&13\\
$n_6$&30&20&24\\
$n_7$&9&6&7\\
$n_8$&27&18&22\\
$n_9$&27&18&22\\
$n_{10}$&30&20&24\\
\hline\end{tabular}
\end{center}
\vspace{-0.5cm}
\end{table}
Here, we discuss the data precision improvement that can be realized with the proposed scheduling utilizing the imprecise computation model with respect to data arrival rate $\lambda$ in the range [1, 2] in increments of 0.1. $\lambda$ can be calculated by dividing the total computation time for the {\it mandatory} and {\it optional} parts by the computation time for the {\it mandatory} part, where only computation time for the {\it optional} part can be varied. The data precision is the percentage value obtained by dividing the total processing time by the requested processing time. 

 We conducted a simulation of the task graph shown in Figure \ref{fig:dsg_sample} using the computation time matrix shown in \textbf{Table} \ref{tab:comp_time2}. Note that {\it CCR} was fixed at 1. In the schedule results obtained using HVLB\_CC, the schedule holes for $n_2$, $n_5$, and $n_8$ were found to be 9, 5, and 12, respectively, by using Eqs. \ref{eq:condition1} and \ref{eq:condition2}. In Figure \ref{fig:dsg_sample}, it has been assumed that $n_2$, whose arrival rate can be increased, is processing the receiving data streams from external communications and changing the data format to be used in the DSMS. $n_5$ is assumed as matching processing with map data from $n_3$ and external-vehicles information from $n_2$. Thus, we changed $n_2$ and $n_5$ into imprecise computation models and input data arrival rates. We conducted two sets of experiments: one set for the case in which imprecise computation models are not allowed (i.e., the original version of HVLB\_CC) and another set for the imprecise computation case (i.e., for the alternative version, HVLB\_CC\_IC). The simulation results were analyzed as follows.

\textbf{Figure} \ref{fig:imprecise} shows a comparison between the original and alternative versions of HVLB\_CC. With HVLB\_CC\_IC, the data precisions of $n_2$ and $n_5$ are 100 \% until the input data arrival rates become 1.5 and 1.4, respectively, and their precisions start to decrease at input data arrival rates of 1.6 and 1.5, respectively. This is because the available schedule holes for $n_2$ and $n_5$ are 9 and 5; thus, when the requested computation time increases beyond the available schedule hole, each task starts to terminate the execution of {\it optional} parts. In contrast, the data precisions of $n_2$ and $n_5$ decrease from 1 in HVLB\_CC without imprecise computations are not allowed, because HVLB\_CC without imprecise computations does not exploit any schedule hole, and each task terminates execution after finishing its {\it mandatory} part. Compared to HVLB\_CC without imprecise computations, HVLB\_CC with the imprecise computation model exhibits the greatest improvement in terms of data precision. 

\begin{table*}[!h]
\centering
\newcolumntype{X}{>{\centering\arraybackslash}p{1.5cm}} 
\newcolumntype{Y}{>{\centering\arraybackslash}p{1.5cm}} 
\renewcommand{\arraystretch}{1.12}
\caption{HVLB\_CC vs Previous Work}
\label{tab:comparison}
%\begin{tabular}{p{3cm}|p{2.8cm}p{2.8cm}p{2.5cm}p{2.5cm}}\hline
%\footnotesize
\scriptsize
%\tiny
\begin{tabular}{p{3.8cm}|XXYYYYY}\hline
  & Heterogeneity of processors & Heterogeneity of networks & Communication contention & Accurate time predictability & Load balance over resources & Schedulability of stream processing & Varying data input arrival rate \\\hline
HEFT, CPOP \cite{list_advantage}  & \xmark & & & & & & \\
Automotive DSMS \cite{crts,aedsms} & & & & & & \xmark & \xmark \\
Clustering-based \cite{clus1,clus2,clus3,clus4,clus5} & \xmark & & & & \xmark & & \\
Comm. contention-based \cite{comm_con,comm_con2} & & & \xmark & & \xmark & \xmark & \\
APN-based \cite{comm_con1} & \xmark & & \xmark & \xmark & & & \\
HSV\_CC \cite{hsv_cc} & \xmark & \xmark & \xmark & \xmark & & & \\
HVLB\_CC & \xmark & \xmark & \xmark & \xmark & \xmark & \xmark & \xmark \\\hline
\end{tabular}
\end{table*}

\section{Related Work}\label{sec:related_work2}
In the automotive field, DSMSs have been researched for applications in automotive embedded systems in order to reduce data processing complexity and software development costs. In a previous study \cite{aedsms}, a scheduling method that uses the earliest deadline first for an automotive embedded DSMS was implemented on each distributed processor. In \cite{crts}, distributed stream processing on single- and multi-core processors was investigated to process numerous streams of data from external communications more efficiently. However, these methods assume that an optimal distribution pattern for multiple operators (tasks) is already known; thus, the distributions of operators and messages on multiple processors and networks have not previously been described in detail. StreamCar \cite{streamcar} is a stream processing platform that facilitates the combination of sensor data. However, this method does not consider strict real-time constraints, which are important in automotive systems, and does not include a scheduling method. 

Most DAG scheduling algorithms are based on list scheduling which maintains a list of tasks in a DAG according to their priorities. This process has two phases: (i) a task prioritizing phase to assign a priority to each task and to queue the tasks according to their priorities and (ii) a processor selection phase to select a task based on its order of priority and to assign the selected task to a suitable processor.

The heterogeneous earliest-finish-time (HEFT) algorithm \cite{list_advantage} calculates the {\it upward rank} value for each task, and the task with the highest {\it upward rank} value is selected first by traversing the graph upwards. The critical-path-on-a-processor (CPOP) algorithm \cite{list_advantage} uses both {\it upward} and {\it downward rank} values, and these values are used to prioritize each task. This algorithm is similar to the HEFT algorithm, but all of the tasks on critical paths are assigned to a single processor that can minimize their execution times. These algorithms select processors for the tasks according to their {\it EFT} values, in which the computation and communication times are calculated as average values. They primarily concentrate on task scheduling to minimize the schedule length without considering message scheduling. Furthermore, they assume simple system models in which all of the processors are fully connected and have concurrent inter-processor communication, even though communication contention can influence the execution time in parallel processing \cite{comm_con}. Therefore, it is difficult to obtain accurate schedule results with these algorithms, especially in automotive embedded systems in which time accuracy is strictly required.

Clustering-based scheduling algorithms \cite{clus1,clus2,clus3,clus4,clus5} primarily target networked on-chip multiprocessor architectures that are generally intended for unbounded numbers of processors. Tasks that are clustered according to some criteria in the DAG are assigned to multiple processors. However, these clustering-based scheduling algorithms commonly use dynamic task allocation methods. When a task graph arrives at the system, it must first enter a {\it global waiting queue} and will subsequently be moved to a {\it local waiting queue} on a processor. This dynamic method is difficult to apply to automotive embedded systems due to the potential for unpredictable communication times. Thus, time predictability of stream processing cannot be guaranteed. Moreover, these algorithms assume simple system models in which communication contention is not considered because particular communication links between two processors are assumed to be prepared, which indicates that communication contention can be ignored. To obtain accurate and efficient scheduling, both task and message scheduling must be considered simultaneously. Therefore, the main weaknesses of clustering-based algorithms are high-complexity and unpredictable timing constraints because they primarily target unlimited numbers of processors. 

Some studies consider the communication contention problem in list scheduling algorithms. Sinnen et al. proposed a communication contention-awareness scheduling algorithm based on list scheduling \cite{comm_con,comm_con2}, and they showed that this algorithm could produce significantly more accurate scheduling than could a contention-unaware algorithm in a hetegeneous computing environment in which all processors had the different processing abilities. Their algorithms addressed contention by providing appropriate communication routing between processors. Tang et al. proposed a list scheduling method with contention on communication links in heterogeneous computing environments with an arbitrary processor network (APN) \cite{comm_con1}. They concentrated on a network topology that consisted of processors connected by the APN, in which constant $p(p-1)/2$ ($p$ represents the number of processors) communication routes between processors existed. For in-vehicle networks, several types of network protocols, such as CAN and MOST, are inter-connected via gateways. Thus, there are multiple routes between processors, and different numbers of communication routes exist between them. Moreover, strict timing constraints cannot be guaranteed in existing scheduling algorithms with communication contention because they do not consider the synchronization of tasks and messages and they assume that communication delays are possible. Thus, it is difficult to measure accurate end-to-end worst-case execution times (WCETs), which are the most important factor in automotive fields. 

Xie et al. \cite{hsv_cc} proposed HSV\_CC based on list scheduling in consideration of communication contention problems in heterogeneous automotive embedded systems. They assumed that the number of communication routes between different processors could differ and selected the most appropriate communication route. Moreover, they could provide task and message synchronization, which is important for measuring accurate end-to-end WCET. However, it is difficult to apply HSV\_CC to an automotive DSMS for the following reasons. First, \boldred{it can cause inefficient resource allocation in which tasks and messages are likely to be assigned to specific processors and network links}. In the worst case, some processors and links are poorly utilized even though other processors and links have sufficient idle time slots. Second, \boldred{HSV\_CC scheduling was designed to schedule DAGs. Thus, it is possible that some stream processing applications for automotive DSMSs may not be scheduled appropriately.} Third, HSV\_CC, which assumes that the processing time of each task is constant, is inappropriate for automotive DSMSs. In stream processing, the requested processing time for a task can vary. For example, the amount of input data for map matching processing using stored map data and the external vehicle information acquired by external communications can vary over time. This type of processing is difficult to be scheduled efficiently with static HSV\_CC. Static list scheduling algorithms such as HSV\_CC or HEFT can be applied to dynamic scheduling to handle tasks whose requested processing time can vary. However, a dynamic scheduling method is not desired in automotive DSMSs due to strict time predictability and resource restrictions.

\textbf{Table} \ref{tab:comparison} shows a comparison between HVLB\_CC and previous methods. HVLB\_CC can be applied to parallel and distributed environments consisting of heterogeneous processors connected to in-vehicle networks that contain different technologies with different bandwidths. To predict execution times for stream processing accurately, the proposed algorithm schedules tasks and messages simultaneously in consideration of the communication contention constraint. For application in automotive DSMSs, HVLB\_CC considers LB, and can deal with SPGs. Furthermore, the alternative version of the proposed algorithm can handle tasks with varying input data arrival rates by applying an imprecise computation model.

\section{Conclusion}\label{sec:conclusion2}
This paper has presented heuristic scheduling algorithms based on static-list scheduling for automotive DSMSs. \boldred{We have addressed three issues in applying the existing static list scheduling algorithm called HSV\_CC, which is proposed for heterogeneous embedded systems, to stream processing distribution: (i) previous task and message schedules can lead to less efficient resource usages in a stream processing scenario, (ii) the conventional method to determine the task scheduling order may not be best suited to deal with stream processing graphs, and (iii) tasks with time-varying computational requirements are needed to be scheduled efficiently.} 

To address (i), we proposed the HVLB\_CC (A) algorithm, which considers LB in addition to the parameters considered by the HSV\_CC algorithm. We proposed HVLB\_CC (B) to address issue (ii). HVLB\_CC (B) can deal with stream processing task graphs and more various DAGs to prevent assigning higher priorities to successor tasks. To schedule tasks more efficiently with various computation times (iii), HVLB\_CC\_IC utilizes schedule holes left in processors. These idle time slots can be used to execute for {\it optional} parts and thereby generate more precise result by applying imprecise computation model. Our experimental results demonstrate that the proposed algorithms achieve minimum schedule lengths, accuracies, and LB values significantly better than those obtainable by using the existing HSV\_CC algorithm. In addition, the proposed HVLB\_CC (B) algorithm can schedule more varied task graphs (including SPG) without reducing performance. Furthermore, using imprecise computation models, HVLB\_CC\_IC obtains higher precision data than HVLB\_CC can without imprecise computation models. Therefore it is possible to improve the accuracy of applications that use V2V communication.

In the future, we plan to extend the proposed algorithm so that it can be applied to task graphs with multiple periods and deadlines. In an automotive DSMS, sensor data from different sources (e.g., radar, camera, and external communication) generally arrive to the system at different periods and exit tasks that produce the final processed sensor data for applications should be able to have different deadlines.
Furthermore, DSMS mechanisms such as data buffering, window aggregation, and operator state should be considered. To experimentally validate the proposed scheduling algorithm, we also plan to conduct experiments using data from real-world data sets and analyze the experimental results.

\section*{Acknowledgments}\label{sec:ack}
This work was supported by JSPS KAKENHI Grant Number 15H05305.



\bibliographystyle{elsarticle-num} 
\bibliography{myBibDB}

\begin{thebibliography}{10}
\expandafter\ifx\csname url\endcsname\relax
  \def\url#1{\texttt{#1}}\fi
\expandafter\ifx\csname urlprefix\endcsname\relax\def\urlprefix{URL }\fi
\expandafter\ifx\csname href\endcsname\relax
  \def\href#1#2{#2} \def\path#1{#1}\fi

\bibitem{its}
{I}ntelligent {T}ransport {S}ystems ({ITS}); {V2X} {A}pplications; {P}art 3:
  {L}ongitudinal {C}ollision {R}isk {W}arning ({LCRW}) {A}pplication
  {R}equirements {S}pecification, ETSI TS 101 539-3, 2013.

\bibitem{70ecu}
A.~Lara, {F}rom {C}omplex {M}echanical {S}ystem to {C}omplex {E}lectronic
  {S}ystem: the {C}ase of {A}utomobiles, in: Automotive Technology and
  Management, Vol.~14, 2014, pp. 65--81.

\bibitem{diff_ecu}
{D}ifferent {T}ypes of {M}icrocontrollers are used in {A}utomobile
  {A}pplications, Available from\
  \url{https://www.elprocus.com/different-microcontrollers-used-in-automobiles}.

\bibitem{in_vehicle_networks}
S.~Seo, J.~Kim, S.~Hwang, K.~Kwon, J.~Jeon, {A} reliable gateway for in-vehicle
  networks based on {LIN}, {CAN}, and {F}lex{R}ay, ACM Trans. Embed. Comput.
  Syst. 11~(1).

\bibitem{crts}
J.~Rho, T.~Azumi, H.~Oyama, K.~Sato, N.~Nishio, {D}istributed {P}rocessing for
  {A}utomotive {D}ata {S}tream {M}anagement {S}ystem on {M}ixed {S}ingle- and
  {M}ulti-core {P}rocessors, {ACM SIGBED} Rev. 13~(3) (2016) 15--22.

\bibitem{aedsms}
A.~Yamaguchi, Y.~Nakamoto, K.~Sato, I.~Yoshiharu, Y.~Watanabe, S.~Honda,
  H.~Takada, {AEDSMS}: {A}utomotive {E}mbedded {D}ata {S}tream {M}anagement
  {S}ystem, in: Proceedings of the 31st International Conference on Data
  Engineering ({ICDE}), 2015, pp. 1292--1303.

\bibitem{streamcar}
A.~Bolles, H.-J. Appelrath, D.~Geesen, M.~Grawunder, M.~Hannibal, J.~Jacobi,
  F.~Koster, S.~Nicklas, D., {S}tream{C}ars: {A} {N}ew {F}lexible
  {A}rchitecture for {D}river {A}ssistance {S}ystems, in: Proceedings of IEEE
  Intelligent Vehicles Symposium (IV), 2012, pp. 252--257.

\bibitem{borealis}
D.~J. Abadi, Y.~Ahmad, M.~Balazinska, U.~Cetintemel, M.~Cherniack, J.-H. Hwang,
  W.~Lindner, A.~S. Maskey, A.~Rasin, E.~Ryvkina, N.~Tatbul, Y.~Xing,
  S.~Zdonik, {T}he {D}esign of the {B}orealis {S}tream {P}rocessing {E}ngine,
  in: Proceedings of the 2nd Biennial Conference on Innovative Data Systems
  Research (CIDR), 2005, pp. 277--289.

\bibitem{aurora}
D.~J. Abadi, D.~Carney, U.~\c{C}etintemel, M.~Cherniack, C.~Convey, S.~Lee,
  M.~Stonebraker, N.~Tatbul, S.~Zdonik, {A}urora: {A} {N}ew {M}odel and
  {A}rchitecture for {D}ata {S}tream {M}anagement, The VLDB Journal 12~(2)
  (2003) 120--139.

\bibitem{telegraph}
S.~Chandrasekaran, O.~Cooper, A.~Deshpande, M.~J. Franklin, J.~M. Hellerstein,
  W.~Hong, S.~Krishnamurthy, S.~Madden, V.~Raman, F.~Reiss, M.~Shah,
  Telegraph{CQ}: {C}ontinuous {D}ataflow {P}rocessing for an {U}ncertain
  {W}orld, in: Proceedings of the 1st Biennial Conference on Innovative Data
  Systems Research (CIDR), 2003.

\bibitem{stream}
D.~Arvind, A.~Arasu, B.~Babcock, S.~Babu, M.~Datar, K.~Ito, I.~Nishizawa,
  J.~Rosenstein, J.~Widom, {STREAM}: {T}he {S}tanford {S}tream {D}ata {M}anager
  (demonstration description), in: Proceedings of the {ACM} {SIGMOD}
  International Conference on Management of data, Vol.~26, 2003, p. 665.

\bibitem{multicore_dsms}
A.~Safaei, A.~Sharifrazavian, M.~Sharifi, M.~Haghjoo, {D}ynamic routing of data
  stream tuples among parallel query plan running on multi-core processors,
  Distrib. Parallel Databases 30~(2) (2012) 145--176.

\bibitem{multicore_dsms2}
S.~Das, S.~Antony, D.~Agrawal, A.~Abbadi, {T}hread {C}ooperation in {M}ulticore
  {A}rchitectures for {F}requency {C}ounting over {M}ultiple {D}ata {S}treams,
  Proceedings of VLDB Endow. 2~(1) (2009) 217--228.

\bibitem{multicore_dsms3}
Y.~Zhou, B.~Ooi, K.-L. Tan, J.~Wu, {E}fficient {D}ynamic {O}perator {P}lacement
  in a {L}ocally {D}istributed {C}ontinuous {Q}uery {S}ystem, in: Proceedings
  of the 14th International Conference on Cooperative Information Systems, Vol.
  4275, 2006, pp. 54--71.

\bibitem{hetero_computing1}
M.~Muthucumaru, B.~Tracy~D., J.~Howard, {H}eterogeneous distributed computing,
  in: Proceedings of {E}ncyclopedia of {E}lectrical and {E}lectronics
  Engineering, Vol.~8, 1999, pp. 679--690.

\bibitem{hetero_computing2}
F.~Dror~G., R.~Larry, S.~Kenneth~C., W.~Parkson, {Theory} and {P}ractice in
  {P}arallel {J}ob {S}cheduling, in: {B}ook of {J}ob Scheduling {S}trategies
  for {P}arallel {P}rocessing, Vol. 1291, 1997, pp. 1--34.

\bibitem{list1}
B.~Rashmi, D.~Agrawal, {I}mproving scheduling of tasks in a heterogeneous
  environment, IEEE Transactions on Parallel and Distributed Systems 15~(2)
  (2004) 107--118.

\bibitem{list2}
B.~Sanjeev, C.~S. Prashanth, {S}cheduling directed a-cyclic task graphs on
  heterogeneous network of workstations to minimize schedule length, in:
  Proceedings of the IEEE International Conference on Parallel Processing
  Workshops, 2003, pp. 97--103.

\bibitem{list3}
B.~Savina, K.~Padam, S.~Kuldip, {D}ealing with heterogeneity through limited
  duplication for scheduling precedence constrained task graphs, J. Parallel
  Distrib. Comput. 65~(4) (2005) 479--491.

\bibitem{list4}
G.~Apostolos, Y.~Tao, {A} comparison of clustering heuristics for scheduling
  directed acyclic graphs on multiprocessors, J. Parallel Distrib. Comput.
  16~(4) (1992) 276 -- 291.

\bibitem{list5}
T.~Hagras, J.~Janecek, A high performance, low complexity algorithm for
  compile-time task scheduling in heterogeneous systems, Parallel Computing
  31~(7) (2005) 653 -- 670, heterogeneous Computing.

\bibitem{clus5}
S.~C. Kim, S.~Lee, J.~Hahm, {P}ush-{P}ull: {D}eterministic {S}earch-{B}ased
  {DAG} {S}cheduling for {H}eterogeneous {C}luster {S}ystem, IEEE Transactions
  on Parallel and Distributed Systems 18~(11) (2007) 1489--1502.

\bibitem{list7}
T.~Xiaoyong, L.~Kenli, L.~Renfa, V.~Bharadwaj, {R}eliability-aware scheduling
  strategy for heterogeneous distributed computing systems, J. Parallel
  Distrib. Comput. 70~(9) (2010) 941--952.

\bibitem{list_advantage}
H.~Topcuouglu, S.~Hariri, M.-Y. Wu, {P}erformance-{E}ffective and
  {L}ow-{C}omplexity {T}ask {S}cheduling for {H}eterogeneous {C}omputing, IEEE
  Transactions on Parallel and Distributed Systems 13~(3) (2002) 260--274.

\bibitem{hsv_cc}
X.~Guoqi, L.~Renfa, L.~Keqin, {H}eterogeneity-driven end-to-end synchronized
  scheduling for precedence constrained tasks and messages on networked
  embedded systems, J. Parallel Distrib. Comput. 83~(C) (2015) 1--12.

\bibitem{aedsms3}
S.~Katsunuma, S.~Honda, K.~Sato, Y.~Watanabe, Y.~Nakamoto, H.~Takada,
  {R}eal-{T}ime-{A}ware {E}mbedded {DSMS} {A}pplicable to {A}dvanced {D}river
  {A}ssistance {S}ystems, in: Proceedings of the IEEE 33rd International
  Symposium on Reliable Distributed Systems Workshops (SRDSW), 2014, pp.
  106--111.

\bibitem{aedsms2}
M.~Yamada, K.~Sato, H.~Takada, {I}mplementation and {E}valuation of {D}ata
  {M}anagement {M}ethods for {V}ehicle {C}ontrol {S}ystems, in: Proceedings of
  the IEEE Vehicular Technology Conference (VTC Fall), 2011, pp. 1--5.

\bibitem{clus1}
T.~Yang, A.~Gerasoulis, {DSC}: scheduling parallel tasks on an unbounded number
  of processors, IEEE Transactions on Parallel and Distributed Systems 5~(9)
  (1994) 951--967.

\bibitem{clus2}
S.~Kim, J.~C. Browne, {A} general approach to mapping of parallel computation
  upon multiprocessor architectures, in: Proceedings of the International
  Conference on Parallel Processing, Vol.~3, 1988, pp. 1--8.

\bibitem{clus3}
J.-C. Liou, M.~A. Palis, {A}n {E}fficient {T}ask {C}lustering {H}euristic for
  {S}cheduling {DAG}s on {M}ultiprocessors, in: Proceedings of Workshop on
  Resource Management, Symposium of Parallel and Distributed Processing, 1996,
  pp. 152--156.

\bibitem{clus4}
G.~L. Stavrinides, H.~D. Karatza, {S}cheduling {R}eal-time {DAG}s in
  {H}eterogeneous {C}lusters by {C}ombining {I}mprecise {C}omputations and
  {B}in {P}acking {T}echniques for the {E}xploitation of {S}chedule {H}oles,
  Future Gener. Comput. Syst. 28~(7) (2012) 977--988.

\bibitem{lb}
F.~A. Omara, M.~M. Arafa, {G}enetic algorithms for task scheduling problem, J.
  Parallel Distrib. Comput. 70~(1) (2010) 13--22.

\bibitem{tgff}
R.~P. Dick, D.~L. Rhodes, W.~Wolf, {TGFF}: task graphs for free, in:
  Proceedings of the 6th international workshop on Hardware/software codesign,
  1998, pp. 97--101.

\bibitem{comm_con}
O.~Sinnen, L.~A. Sousa, {C}ommunication contention in task scheduling, IEEE
  Transactions on Parallel and Distributed Systems 16~(6) (2005) 503--515.

\bibitem{comm_con2}
O.~Sinnen, L.~A. Sousa, F.~Sandnes, {T}oward a {R}ealistic {T}ask {S}cheduling
  {M}odel, IEEE Transactions on Parallel and Distributed Systems 17~(3) (2006)
  263--275.

\bibitem{comm_con1}
X.~Tang, K.~Li, D.~Padua, {C}ommunication contention in {APN} list scheduling
  algorithm, in: Proceedings of Science in China Series F, Information
  Sciences, Vol.~52, 2009, pp. 59--69.

\end{thebibliography}


%\begin{thebibliography}{00}


%\bibitem{}
%\bibliography{myBibDB}
%\end{thebibliography}

\clearpage
\end{document}